%% file: ANA-BPHY-2018-05-PAPER.tex
\documentclass[PAPER, atlasdraft=false, UKenglish, cernpreprint, texmf, orcidlogo]{atlasdoc}
\usepackage{atlaspackage}
\usepackage{atlasbiblatex}

\usepackage{atlasphysics}

\graphicspath{{logos/}{figures/}}

\usepackage{ANA-BPHY-2018-05-PAPER-defs}

\AtlasTitle{A search for lepton-flavour violating $\tau \to 3\mu$ decays with the ATLAS detector}

\AtlasAbstract{%
A search for charged lepton flavour violation in $\tau \to 3\mu$ decays is performed in $pp$
collisions at a centre-of-mass energy of 13~TeV using ATLAS data collected between 2016 and 2018,
corresponding to an integrated luminosity of 137~$\text{fb}^{-1}$. The search is optimised for the electroweak $W \to \tau\nu$ production channel, due to its higher trigger efficiency, but also considers potential signal events from heavy-flavour production. Data are collected using two-muon and three-muon triggers and a multivariate
analysis is used to separate the signal from the background. An unbinned likelihood fit
is then performed to the resulting three-muon invariant mass spectrum and the data are found to be
compatible with the background-only hypothesis. The observed (expected) limit on the branching ratio $B(\tau\to3\mu)$ is
found to be $8.7\times 10^{-8}$ ($7.5\times 10^{-8}$) at 90\% CL.  }

\AtlasRefCode{BPHY-2018-05}

\PreprintIdNumber{CERN-EP-2026-059}

\AtlasJournal{Phys.\ Lett.\ B.}
\AtlasJournalRef{\PLB 880 (2026) 140748}
\AtlasDOI{10.1016/j.physletb.2026.140748}

\hypersetup{pdftitle={ATLAS document},pdfauthor={The ATLAS Collaboration}}

\begin{document}

\maketitle

\tableofcontents

%
%

%

\section{Introduction}
In the Standard Model (SM), flavour is not a fundamental symmetry and
its violation is observed in both the
neutrino~\autocite{neutrino_oscillations_sk} and quark
sectors~\autocite{ckm1,ckm2}. It is therefore natural to question
whether flavour conservation is violated for charged leptons.
Although neutrino oscillations provide a route for charged Lepton Flavour Violation (cLFV) in the SM, the corresponding branching ratios ($B$) are vanishingly small, of the order
$10^{-55}$~\autocite{decay_tau_3mu_sm}. Any observation of cLFV would
therefore be unambiguous evidence of physics beyond the SM
(BSM). Searches for cLFV have been carried out in the decays of both muons and
$\tau$-leptons, however, the most stringent limits on cLFV for the $\tau$-lepton
are currently about four orders of magnitude
weaker than those for the muon decay channel~\autocite{megii_muegamma,Belle_II}.
Theoretical arguments suggest that new
physics effects would be stronger for the heavier
$\tau$-lepton~\autocite{increased_tau_coupling}. Several BSM
theoretical frameworks provide an enhancement to the rate of cLFV,
with predicted $\tau$-lepton flavour violating $B$ of the order
$10^{-9}$~\autocite{t1_seesaw_clfv_limits,t1_seesaw_clfv_limits2,z_prime_limits,susy_limit,
thdm_limit}, potentially detectable by current experiments.

This paper presents a new search for the cLFV process $\tau^{\pm}
\rightarrow \mu^{\pm}\mu^{\mp}\mu^{\pm}$, henceforth referred to as $\tau \rightarrow 3 \mu$.
The most stringent limits on this process are obtained by the Belle II and LHCb experiments. The
Belle II experiment set an observed (expected) limit of $1.9 \times10^{-8}$ ($1.8
\times10^{-8}$) at $90\%$ confidence level (CL)~\autocite{Belle_II},\footnote{All limits are reported at $90\%$~CL in line with previous results.} using $424\,\ifb$ of $e^+e^-$ collision data with a centre-of-mass energy near the $\Upsilon(4S)$ resonance. Other stringent limits were set using  $e^+e^-$ collision data by the Belle experiment with an observed limit of $2.1 \times10^{-8}$~\autocite{Belle} using $782\,\ifb$ of data, and by the BaBar experiment with an observed (expected) limit
of $3.3 \times10^{-8}$ ($4.0 \times10^{-8}$)~\autocite{Babar} using $468\,\ifb$ of data. Using proton--proton ($pp$) collision data at the Large Hadron Collider (LHC) at $\sqrt{s} = 13\,\TeV$, the LHCb experiment set an observed (expected) limit of $1.9\times10^{-8}$ ($2.0 \times10^{-8}$)~\autocite{lhcb_run2} using $5.4\,\ifb$ of data collected during Run 2, matching the previous most stringent observed limit. The CMS experiment set an observed (expected) limit of $2.9 \times10^{-8}$ ($2.4\times10^{-8}$)~\autocite{CMS-BPH-21-005} using $131\,\ifb$ of data collected during Run 2. The most recent analysis by the ATLAS experiment used $20.3\,\ifb$ of data collected during Run 1 at $\sqrt{s} = 8\,\TeV$ to set an observed (expected) limit of $37.6 \times10^{-8}$ ($39.4 \times10^{-8}$)~\autocite{EXOT-2014-14}.

The search for  $\tau \rightarrow 3 \mu$ presented here uses 137~\ifb of $pp$
collision data at $\sqrt{s} = 13\,\TeV$ collected from 2016--2018 by the ATLAS experiment. This is
the first ATLAS search for this process using Run 2 data, and supersedes the previous
result. At the LHC, $\tau$-leptons are produced in large quantities either through the decays of heavy flavour hadrons,
referred to as the heavy-flavour (HF) signal, or through the decays of $W$
and $Z$ bosons, including $W$ bosons from top-quark decays, referred to as the electroweak (EW) signal.
Although the HF signal, primarily composed of $D_s \to \tau\nu_\tau$ decays, dominates at the LHC,
the $\tau$-leptons produced by the EW signal have on average a larger transverse momentum and are better isolated from hadronic activity. Consequently a large fraction of the HF signal is not recorded by ATLAS due to trigger \pt threshold limitations, and the analysis is instead optimised for the $W \rightarrow \tau{\nu}_\tau$ contribution. The statistical interpretation considers both the full EW and HF $\tau$-lepton production mechanisms, which are henceforth referred to as the signal.

The signature of interest consists of three muons from a secondary vertex, with a three-muon invariant mass
$m_{3\mu}$ consistent with the $\tau$-lepton mass. The data were collected using multi-muon
triggers. Candidate three-muon vertices are then reconstructed. Simulated signals
and sideband data are used in a multivariate analysis (MVA)
to classify events optimally in view of the small expected signal-to-background ratio.
Both the signal and background are parameterised
using functional forms in the $m_{3\mu}$ spectrum. A signal-plus-background fit is performed
simultaneously in all fit categories, allowing the smooth background
to be estimated from the data sidebands while searching for a resonant
signal close to the $\tau$-lepton mass.


%

\section{ATLAS detector}
The ATLAS experiment~\cite{PERF-2007-01} at the LHC is a multipurpose particle detector with a forward--backward symmetric cylindrical geometry and a near \(4\pi\) coverage in solid angle.\footnote{\AtlasCoordFootnote} It consists of an inner tracking detector surrounded by a thin superconducting solenoid providing a \qty{2}{\tesla} axial magnetic field, electromagnetic and hadronic calorimeters, and a muon spectrometer.

The inner detector (ID) covers the pseudorapidity range \(|\eta| < 2.5\). It consists of silicon pixel, silicon microstrip, and transition radiation tracking detectors, with the addition for Run 2 of an inner pixel layer known as the insertable B-layer (IBL).
Lead/liquid-argon (LAr) sampling calorimeters provide electromagnetic (EM) energy measurements
with high granularity within the region \(|\eta|< 3.2\). A steel/scintillator-tile hadronic calorimeter covers the central pseudorapidity range (\(|\eta| < 1.7\)). The endcap and forward regions are instrumented with LAr calorimeters for EM and hadronic energy measurements up to \(|\eta| = 4.9\).

The muon spectrometer (MS) surrounds the calorimeters and is based on three large superconducting air-core toroidal magnets with eight coils each. The field integral of the toroids ranges between \num{2.0} and \qty{6.0}{\tesla\metre} across most of the detector. The muon spectrometer includes a system of precision tracking chambers up to \(|\eta| = 2.7\) and fast detectors for triggering up to \(|\eta| = 2.4\).

The luminosity is measured mainly by the LUCID--2~\cite{LUCID2} detector that is located close to the beampipe.
A two-level trigger system was used to select events~\cite{TRIG-2016-01}. The first-level trigger is implemented in hardware and used a subset of the detector information to accept events at a rate close to \qty{100}{\kHz}. This is followed by a software-based trigger that reduced the accepted rate of complete events to \qty{1.25}{\kHz} on average
depending on the data-taking conditions.

A software suite~\cite{SOFT-2022-02} is used in data simulation, in the reconstruction and analysis of real and simulated data, in detector operations, and in the trigger and data acquisition systems of the experiment.


\section{Data and simulated samples}
The data were taken during $pp$ collisions by ATLAS, during the years 2016--2018
of the Run 2 LHC data-taking period. This dataset corresponds to an integrated luminosity of $137\,\ifb$~\autocite{DAPR-2021-01} at $\sqrt{s} = 13$ \TeV. Only events recorded when LHC beams were
stable and all ATLAS detector systems were operational are considered~\autocite{DAPR-2018-01}.
Data collected in 2015 with an additional luminosity of $3\,\ifb$ are omitted, as the dedicated triggers described in Section~\ref{sec:selection} were not yet fully operational.

For the EW signal, three sources of $\tau$-lepton production are modelled: $W \rightarrow
\tau{\nu}_\tau$, $Z \rightarrow \tau \tau$ and $t \bar{t} \rightarrow \tau X$. The \POWHEGBOX MC event generator~\cite{Nason:2004rx,Frixione:2007vw,Alioli:2010xd,Alioli:2008gx,Frixione:2007nw} was used to model these signal samples at next-to-leading order (NLO) in QCD. Events are interfaced with the \PYTHIA[8.245]~\autocite{Sjostrand:2014zea} generator to model the parton shower,
hadronisation, underlying event, and particle decays. The $W
\rightarrow \tau{\nu}_\tau$ and $Z \rightarrow \tau \tau$ samples used the AZNLO~\autocite{STDM-2012-23}
set of tuned parameters (tune) and the \CT[10nlo] PDF set~\cite{Lai:2010vv}
for the hard-scattering processes, whereas the \CTEQ[6L1] PDF
set~\cite{Pumplin:2002vw} was used for the parton shower. Top-quark pair-production with at least one $\tau$-lepton in the decay ($t \bar{t} \rightarrow \tau X$) used the A14~\autocite{ATL-PHYS-PUB-2014-021}
tune with the \NNPDF[3.0nlo]~\cite{Ball:2014uwa} PDF set for the hard scattering process and the \NNPDF[2.3lo]~\autocite{Ball:2012cx} PDF set for the parton shower. These simulated samples are normalised to
$W,Z$ and $t\bar{t}$ production cross-sections as measured in ATLAS data for $\sqrt{s} = 13$ \TeV~\autocite{STDM-2018-17, TOPQ-2018-26}. Given the current stringent experimental limits on $B(\tau \to 3\mu) < 10^{-7}$, only one of the $\tau$-leptons is allowed to decay into three muons for the $Z$ and $t \bar{t}$ samples. At production level, the $W$ production mode provides $83\%$ of all $\tau$-leptons in the EW signal, followed by the $Z$ production mode ($16\%$) with only minor contributions from the $t\bar{t}$ production mode ($1\%$).

Three HF signal modes of $\tau$-lepton production are modelled: $D_s$ mesons decaying to $\tau$-leptons, with the $D_s$ either being produced directly (prompt) or through $B$-meson decays (non-prompt); and direct $B$-meson to $\tau$-lepton decays. The \PYTHIA8 generator was used to create these signal samples at leading-order QCD with the A14 tune and \NNPDF[2.3lo] PDFs. Decays of heavy-flavour hadrons and $\tau$-leptons are performed by \EVTGEN~\autocite{Lange:2001uf}. The $D_s$ samples are normalised to the production cross-section ($\sigma = 5000 \pm 360 \text{ (stat)} \pm 590 \text{ (syst)}\,\si{\nb}$) measured by ATLAS~\autocite{BPHY-2018-10} at $\sqrt{s} = 13$ \TeV with a branching ratio of $B(D_s \to \tau\nu_\tau)= 5.4 \pm 0.1 \%$~\autocite{Zyla:2020zbs}. The prompt and non-prompt contributions to $D_s$ production are predicted in simulation to be $80\%$ and $20\%$, respectively, although the exact value for this fraction was not found to make a significant difference to the analysis. The direct $B$-meson sample was scaled relative to the total $D_s$ sample using a value of $48\%$ based on Monte Carlo simulation. Furthermore, the branching ratio $B(B \to \tau X) = 3.4 \pm 0.4\%$~\autocite{CMS-BPH-17-004} is taken into account in the normalization of this sample.

The $\tau \to 3 \mu$ decays are modelled isotropically with the
kinematics of the decay products determined via phase space
without polarisation or spin effects taken into account, as there are
multiple possible new physics scenarios with different spin correlations. This is a standard benchmark setting also used in previous analyses~\cite{Belle_II,lhcb_run2,CMS-BPH-21-005}.

All the generated events are passed through the simulation of the ATLAS detector based on \GEANT~\cite{Agostinelli:2002hh, SOFT-2010-01} and reconstructed with the same algorithms as used for data.
The effect of multiple interactions in the same and neighbouring bunch crossings (pile-up) was
modelled by overlaying the simulated hard-scattering event with inelastic \(pp\)
events generated with \PYTHIA[8.186]~\cite{Sjostrand:2007gs} using the \NNPDF[2.3lo] PDF set~\cite{Ball:2012cx} and the A3 tune~\cite{ATL-PHYS-PUB-2016-017}. The simulated events are weighted to reproduce the
distribution of pile-up observed in the data.


%
%
%

\section{Object reconstruction} \label{sec:reconstruction}

Muons are reconstructed from hits in both the ID
and MS sub-detectors, with the track parameters determined using the
combined information~\autocite{MUON-2018-03}. Muons must have a transverse
momentum (\pt) greater than $4$~\GeV, and $|\eta|\leq2.4$. They
must also satisfy the ``Loose'' identification criteria~\autocite{MUON-2018-03}.

To form a three-muon decay candidate, referred to as a triplet, all
events containing at least three muons with a total charge of $Q=\pm1$
are selected. These muons must share the same
interaction origin point. Due to the
relatively long lifetime of the $\tau$-lepton, this interaction vertex may be
displaced on average from the primary vertex (PV),
resulting in a secondary vertex
(SV). The SV is determined using the muon track parameters and
uncertainties to find the optimal intersection point of the three
tracks by minimising the $\chi^2$ of the fit \autocite{sv}. If multiple SV
candidates are found for the same event, the candidate is used that
has the smallest impact parameter of the muon-triplet
momentum with respect to the PV ($a_0$). The muon-triplet momentum is calculated
as the vector sum of the individual muon momenta.
Tracks determined to originate from a SV are refitted with
respect to this vertex and the PV is refitted excluding the tracks
associated with the SV. The triplet direction is defined as the vector
between the PV and SV. This vertexing algorithm is closely related to that
used in the ATLAS $D_s$ production cross-section measurement~\autocite{BPHY-2018-10}.

Hadronic jets are reconstructed using
the \antikt algorithm \autocite{Cacciari:2008gp} as implemented in
FastJet \autocite{Fastjet} with a jet radius parameter $R = 0.4$. The
inputs to this algorithm are particle flow objects
\autocite{PERF-2015-09}, which combine measurements from the ATLAS
inner detector and calorimeters. Only tracks matched to the PV are
used to mitigate pile-up effects. The jets are calibrated to the
electromagnetic scale~\autocite{PERF-2015-09}, with their axis
corrected to point back to the PV. Jets are corrected for pile-up
contributions, energy loss in uninstrumented parts of the
detector and data-to-MC differences. Any jet within $\Delta R =0.2$ of a
muon, or matched with the inner-detector track of a muon
using ghost association~\autocite{jet_overlap}, is removed
if it has fewer than three associated tracks.

The missing transverse momentum
($\vec{p}_{\text{T}}^{\text{miss}}$) (with
magnitude \met)~\autocite{JETM-2020-03} is calculated as the negative vector sum of the
transverse momenta for muons, jets, and additional
contributions. These additional contributions may be calculated either using ID tracks
associated with a vertex but not matched to a jet or muon
or using unmatched clusters of energy
in the calorimeter. The missing transverse momentum calculated using these
contributions is referred to as $\vec{p}_\mathrm{T}^\mathrm{track-miss}$
and $\vec{p}_\mathrm{T}^\mathrm{calo-miss}$, respectively.

All reconstructed simulated objects have their reconstruction and selection
efficiencies and their energy/momentum scale and resolution
corrected to data.


\section{Trigger and event selection} \label{sec:selection}

The data were collected with several two- and
three-muon triggers, with small variations between the
three years of data taking. In each year, a trigger requiring one
muon in the first-level trigger system with $\pt \geq 20\,\GeV$ and a
second muon in the software-based higher-level trigger system with $\pt \geq 6\,\GeV$ provides the highest efficiency
for signal candidates with at least one high-momentum muon. The
intermediate momentum range is covered by triggers that require two
muons in the first-level trigger system with $\pt\geq 10\,\GeV$ for
2016 data, while in 2017 and 2018 a trigger with $\pt\geq 11\,\GeV$
and $\pt\geq 6\,\GeV$ for the leading and subleading muon provides
better efficiency. Finally, a trigger requiring three muons in the
first-level trigger system with $\pt \geq 4\,\GeV$ provides a lower
threshold on muon momenta to maximize the signal acceptance for all years.

A particular challenge for the trigger system are close-by muons from the
decays of high-momentum $\tau$-leptons that
result in overlapping muon trigger regions of interest, reducing
the total trigger efficiency~\autocite{TRIG-2018-01}.
The simulated signal
trigger efficiency is corrected via a data-driven method where the
multi-muon trigger efficiency is calculated as product of individual
muon trigger efficiencies as a function of \pt and $q \times \eta$ (where $q$ is the muon charge), and correction
factors as a function of $\Delta R$. These individual trigger
efficiencies and correction factors are measured in data from
$J/\psi \rightarrow \mu \mu$ events using a tag-and-probe method~\autocite{TRIG-2018-01}.
The nominal results are obtained with the corrected trigger efficiencies and
the full difference to the uncorrected trigger efficiencies is applied as a systematic
uncertainty.

The analysis considers events in an extended three-muon mass range
around the $\tau$-lepton mass, $1500\,\MeV < m_{3\mu} <
2000\,\MeV$, which is the region used for fits. All analysis steps are initially completed by excluding
the range $1700\,\MeV < m_{3\mu} < 1850\,\MeV$ that would contain
$95\%$ of the signal, which is referred to as the signal region. The sidebands are used to model the background,
composed primarily of random combinations of one or two muons with
additional tracks that are misidentified as belonging to muons.
This approach to background modelling was used, as simulations are
unreliable for modelling combinatorial background and
would require an extremely large sample size.

The analysis is optimised to be sensitive to very small signal-to-background
ratios. By comparing expected signal and sideband
background distributions, loose preselections are applied to remove
background-dominated regions while leaving plenty of events to train
an MVA classifier later. To reject poor quality SVs, the SV
$\chi^2$ must be less than $40$ and the transverse SV
impact parameter ($a_0^{\text{xy}}$) of the triplet divided by its uncertainty ($\sigma_{a_0^{\text{xy}}}$) is
required to be $a_0^{\text{xy}}/\sigma_{a_0^{\text{xy}}}<6$.
Triplet candidates with significant close-by activity in the calorimeters
are removed with a selection on
the sum of transverse energy ($E_\text{T}$)
of energy clusters within a cone of $\Delta R = 0.2$ around
the triplet four-momentum~\autocite{MUON-2018-03}, $E_\text{T}^\text{iso} (\Delta R = 0.2)<20\,\GeV$.
Similarly, triplet candidates are required to be isolated from tracks in the ID, excluding
those associated with the muon triplet, by limiting
the scalar sum of the \pt of ID tracks to $\pt^\mathrm{iso} /\pt^\mathrm{triplet} <1.0$, where $\pt^\mathrm{iso}$ is calculated as a sum
over tracks within $\Delta R = \min(0.3, 10\ \GeV/\pt^\mathrm{triplet})$ of the triplet four-momentum for muons with $\pt<50\,\GeV$,
and $\Delta R = 0.2$ otherwise. Selections are also applied
to the refitted di-muon invariant mass for each two-muon
combination within the triplet (both same-sign and opposite-sign charge),
$m_{\mu \mu} > 325\,\MeV$, to reject muons produced
from final state radiation photon conversions and low-mass background. After the preselection $960,000$ events remain in the background sidebands.

Another minor source of background is from muon pairs produced in the decay
of $\omega(782)$ and $\phi(1020)$ mesons
combined with a third muon, or a track misidentified as belonging to a muon, to create an SV mistaken for signal. Events containing an oppositely
charged muon combination with a di-muon invariant mass within $34\,\MeV$ of these meson masses~\autocite{Zyla:2020zbs}
are removed.

The signal-to-background ratio is improved further using a gradient
boosted decision tree (BDT), implemented using XGBoost \autocite{xgboost}. Several deep neural network models were also tried, but the BDT was found to give the best overall performance and classification accuracy. After the application of the
preselection, both the simulated EW signal events
and aforementioned background events from the sidebands in real data are split into two equal halves (based on odd
and even event numbers) and used to train two BDTs, which are then subsequently used to classify the
events in the other sample. This splitting method is a version
of the $k$-fold method with $k=2$
\autocite{decision_trees}, and aims to reduce the bias of the BDT. The values of the
BDT hyperparameters are determined to optimise the model accuracy:
a maximum BDT depth of 8 is used; the minimum child weight, a measure
of sample size at each node, is set to 3; and the hyperparameter $\eta$, related to the amount the tree
weights are adjusted at each iteration, is set to 1. At each stage of node evaluation, $80\%$ of the data, randomly chosen, are used, to prevent over-training. The other hyperparameters are set
to their default values, and the logistic loss function~\autocite{xgboost} is used.

The BDT uses 18 input variables related to the vertex quality, muon isolation, muon triplet kinematics, and correlations between the muon triplet and \met. All variables are defined in Table~\ref{bdttable}, with examples of each type shown in Figure~\ref{bdtplots}. The variables are selected to ensure the BDT cannot reconstruct the triplet mass and sculpt the mass distribution. Certain variables, such as $m_{3 \mu}^{\text{miss}}$ and $\Delta m_{3 \mu}$, were initially optimised for $\tau$-leptons from HF decays. However, as they are found to not compromise the BDT performance, they are kept. While many of the input features contribute to the discrimination power of the BDT, the most important variables are $\Delta\phi(\text{triplet}, \vec{p}_\mathrm{T}^{\text{calo-miss}})$, $|a_0^{xy}/\sigma_{a_0^{xy}}|$ and $\pt^{\text{triplet}}$.

\begin{table}[p]
\caption{List of input variables used in the BDT and their definitions.}
\label{bdttable}
\begin{tabular}{p{0.23\textwidth} | p{0.75\textwidth}}
\hline
Variable                                           & Description \\
\hline
\multicolumn{2}{c}{\textbf{Vertex quality}} \\
\hline
Track $p$-value product  & Product of the three muons refitted ID track $p$-values from the track fit. \\ %
$\chi_{4\text{-track}}^2$ & $\chi^2$-value of the SV with $4$ ID tracks instead of the usual $3$. This variable helps identify events where the triplet contains muons from independent vertices with further charged particles. \\
Vertex $p$-value & The $p$-value associated with the triplet vertex from the SV fit. \\ %
\hline
\multicolumn{2}{c}{\textbf{Muon isolation}} \\
\hline
$E_\mathrm{T}^\mathrm{iso} (\Delta R = 0.2)$ & Muon calorimeter isolation with $\Delta R = 0.2$. \\
$E_\mathrm{T}^\mathrm{iso} (\Delta R = 0.4)$ & Muon calorimeter isolation with $\Delta R = 0.4$. \\
$\pt^\mathrm{iso}/\pt^{\text{triplet}}$ & Muon track isolation with $\Delta R = 0.2$ divided by the triplet \pt. \\
\hline
\multicolumn{2}{c}{\textbf{Triplet related quantities}} \\
\hline
$|a_0^{xy}/\sigma_{a_0^{xy}}|$ & Absolute value of the significance of the transverse SV IP. \\
$L^{xy}/\sigma_{L^{xy}}$ & Significance of the distance in the transverse plane from the PV to the SV projected onto the direction of the vertex candidate's transverse momentum.\\
$\Delta\phi(\text{triplet}, \vec{p}_\mathrm{T}^{\text{track-miss}})$ & Azimuthal angle between the triplet and the track-based \met calculation. \\ %
$\Delta\phi(\text{triplet}, \vec{p}_\mathrm{T}^{\text{calo-miss}})$ & Azimuthal angle between the triplet and the calorimeter-based \met calculation.\\
$\pt^{\text{triplet}}$ & Transverse momentum of the vector sum of the refitted muon momenta \\
$\eta_{\text{triplet}}$ & $\eta$ of the refitted SV of the triplet. \\ %
$m_{\text{coll-jet}}$ & Invariant mass of the closest jet reconstructed within $\Delta R\leq 0.4$ to the muon triplet.   \\

$\left|\vec{p}_\mathrm{T}^\text{triplet}- \vec{p}_\mathrm{T}^\text{jet}\right|$ & Momentum balance between the triplet candidate and the leading \pt hadronic jet, if a jet is reconstructed. \\
$\left|\sum_\text{PV} \vec{p}_\mathrm{T}^{\text{track}}\right|$ & Modulus of the vector sum of the \pt of all tracks with $\pt >1\,\GeV$ associated with the PV of the triplet.\\
$m_{\text{T}}^{\text{track}}$ & The transverse mass of the triplet, calculated using the triplet \pt and the track-based \met: \newline $m_{\text{T}}^{\text{track}} = \sqrt{2 E_{\text{T}}^{\text{track-miss}} p^{\text{triplet}}_{\text{T}}  (1 - \cos(\Delta\phi(\text{triplet}, E_{T}^{\text{track-miss}})))}$. \\

$m_{3 \mu}^{\text{miss}}$ & A measure of the missing energy at the point of the initially decaying particle $m_{3 \mu}^{\text{miss}} = \sqrt{m_{\text{I}}^2 + m_{3 \mu}^2 - 2 E_{\text{I}} E_{\text{triplet}} + 2 p_{\text{I}} p_{\text{triplet}} \cos (\Delta \alpha_{S-P})}$. This variable was optimised for the HF signal from a $D_s$ decay, so it uses $m_{\text{I}} = m_{3 \mu} + m_{D_s} - m_{\tau}$, where $E_{\text{I}}$ is the total energy at the PV, $\vec{p_I} = \frac{m_{\text{I}}}{m_{3 \mu}} p_{\text{triplet}} \hat{S}$ is the scaled momentum of the initially decaying particle with magnitude $p_{\text{I}}$, $p_{\text{triplet}}$ is the total momentum of the muons triplet, $\hat{S}$ is the unit vector pointing from the PV to SV, and $\Delta \alpha_{S-P}$ is the angle between $\vec{p_{\text{triplet}}}$ and $\hat{S}$.   \\

$\Delta m_{3 \mu}$ & A measure of the maximum possible missing energy at the point of the initially decaying particle, effectively the invariant mass of the initially decaying particle, minus the mass of the triplet: $\Delta m_{3 \mu} = \sqrt{2(p^{\text{triplet}}_{\text{T}} \sin( \Delta \alpha_{S-P}))^2 + m_{3 \mu}^2} - m_{3 \mu}$.\\
\hline
\end{tabular}
\end{table}

\begin{figure}
\centering
\subfloat[]{
\includegraphics[width=0.45\linewidth]{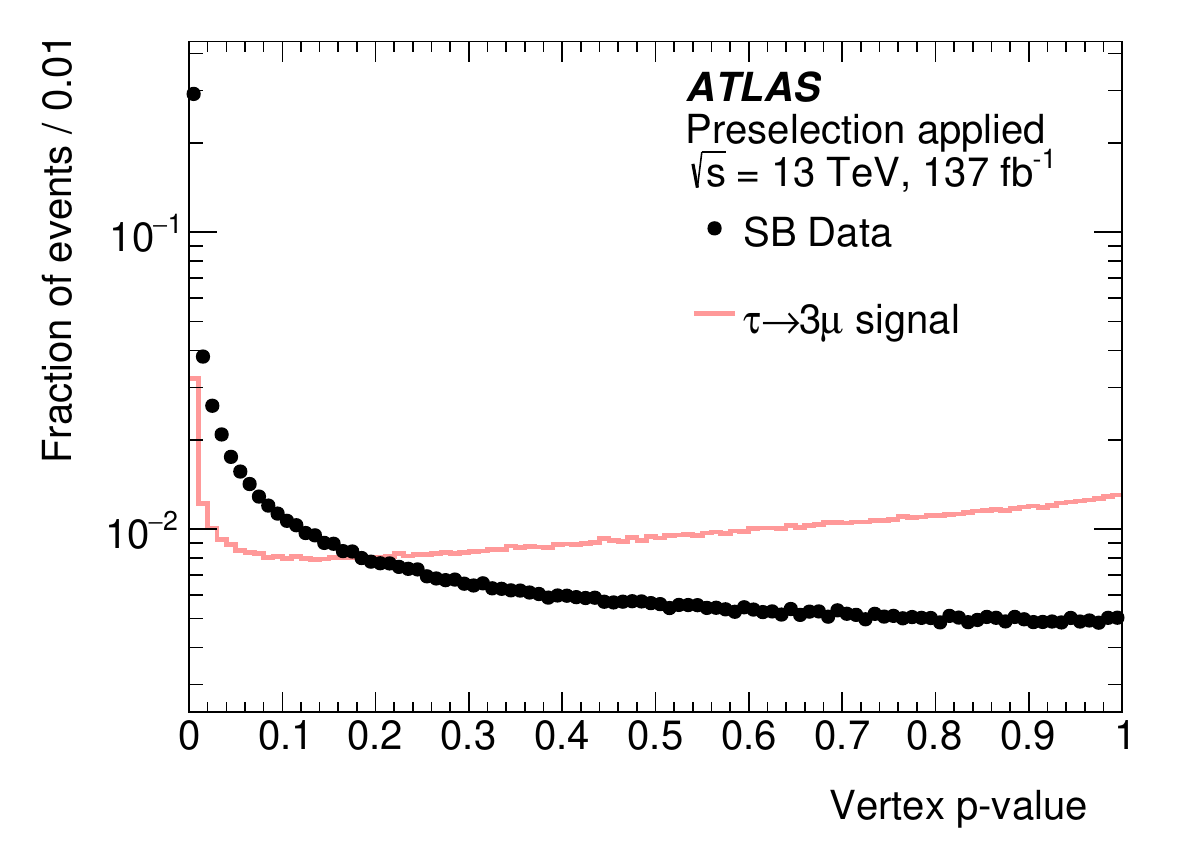}
}
\subfloat[]{
\includegraphics[width=0.45\linewidth]{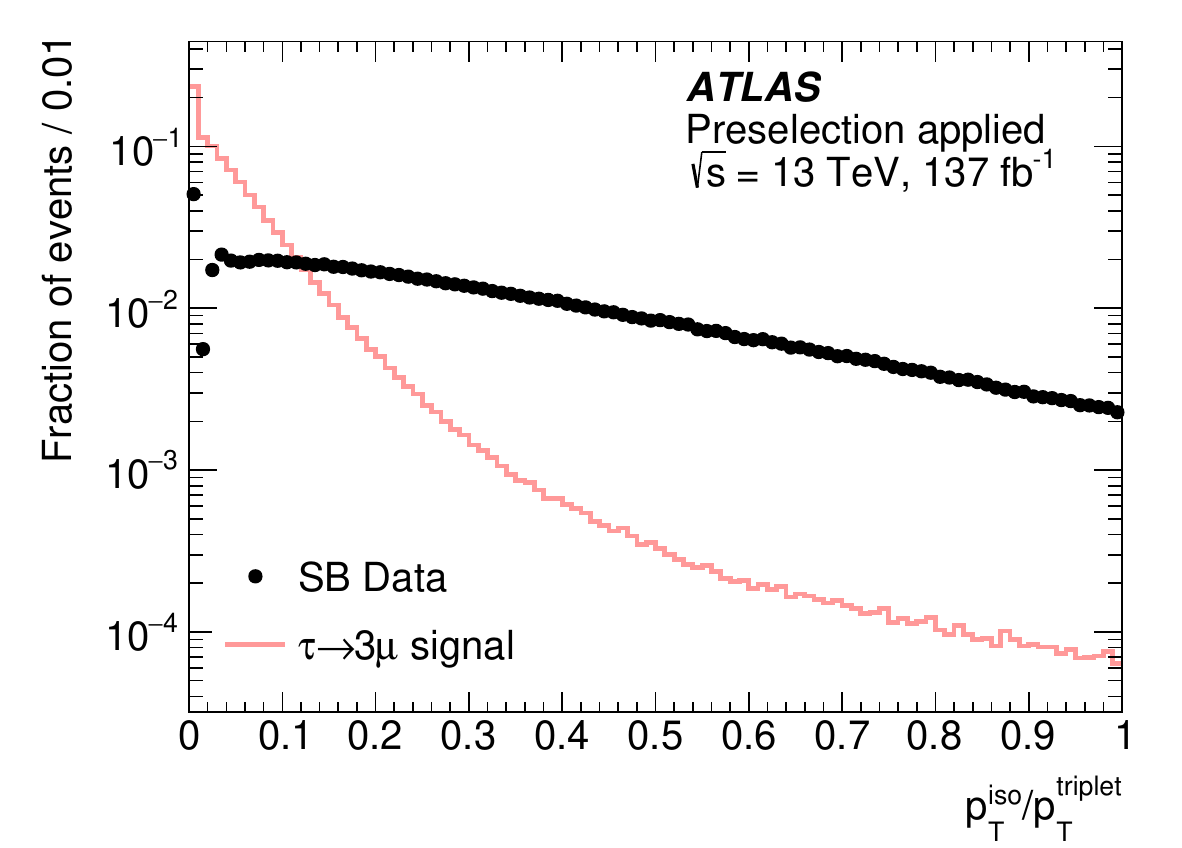}
} \hfill
\subfloat[]{
\includegraphics[width=0.45\linewidth]{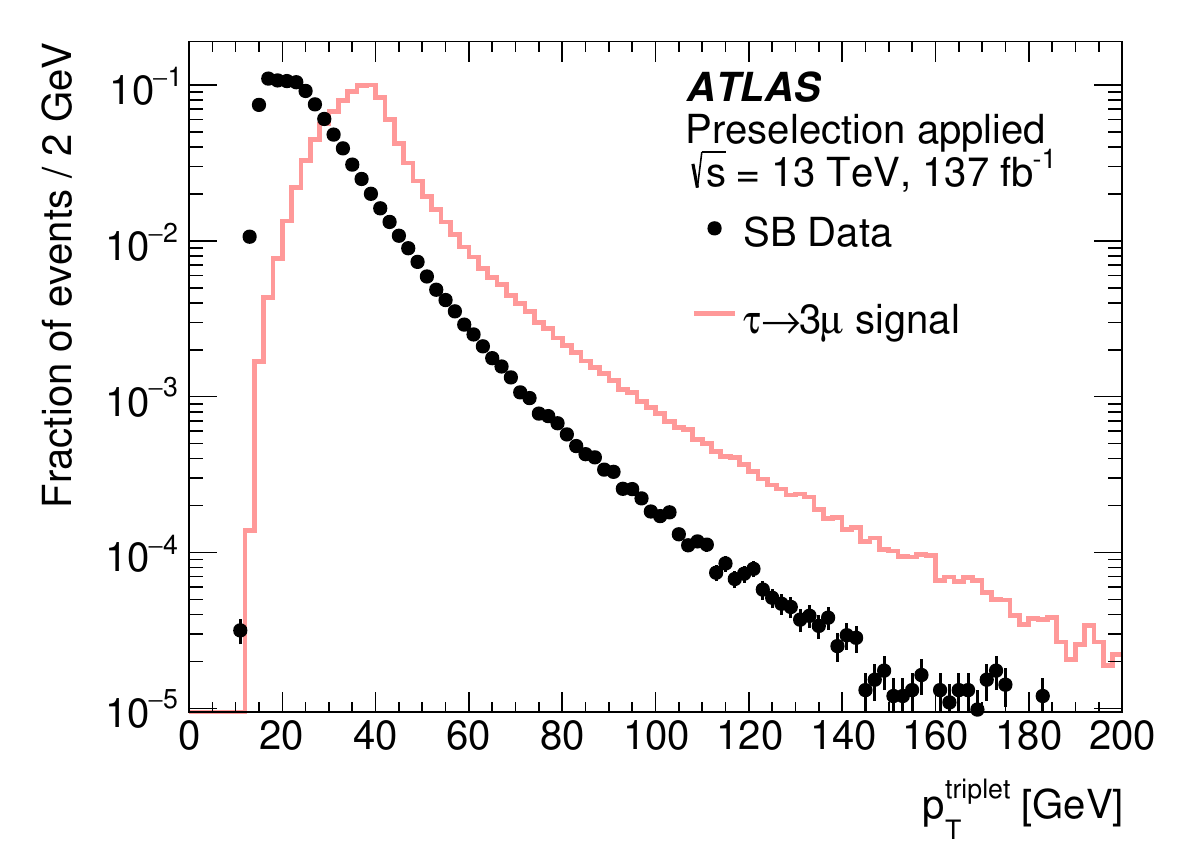}
}
\subfloat[]{
\includegraphics[width=0.45\linewidth]{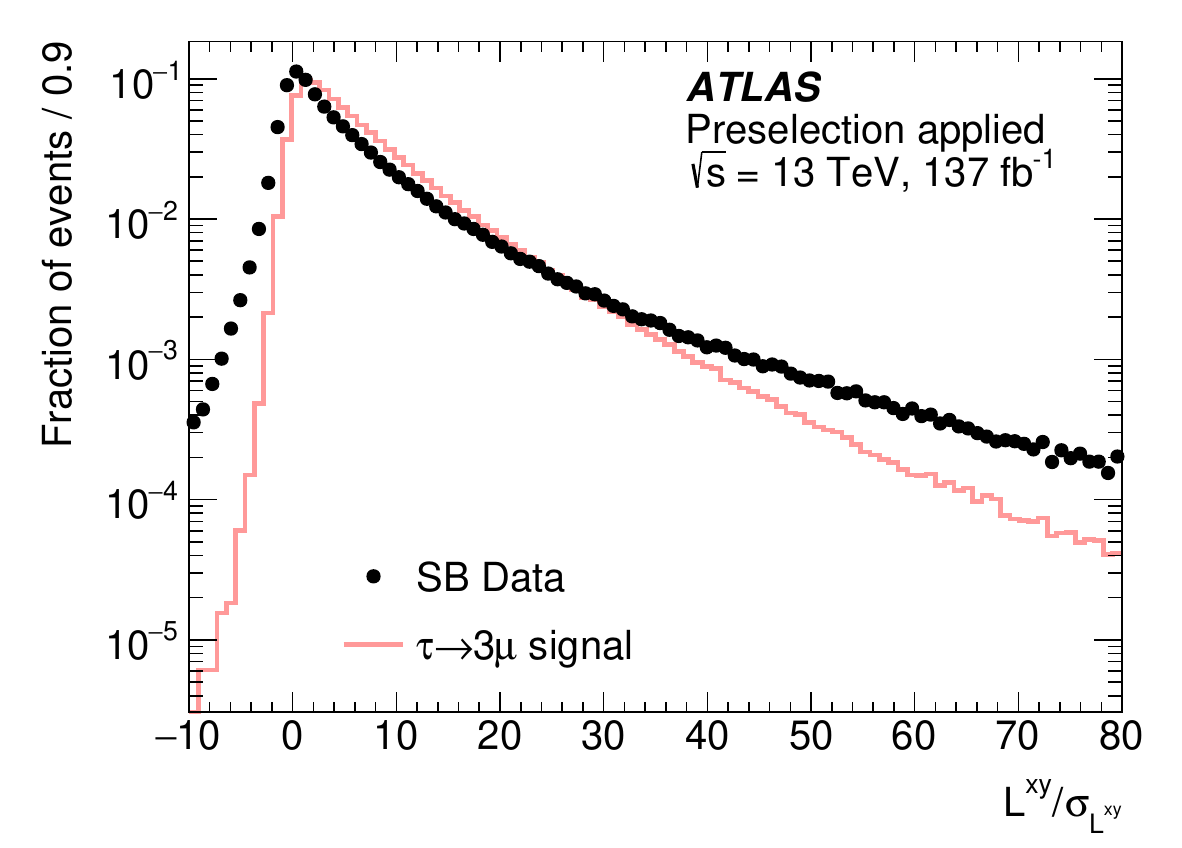}
}  \hfill
\subfloat[]{
\includegraphics[width=0.45\linewidth]{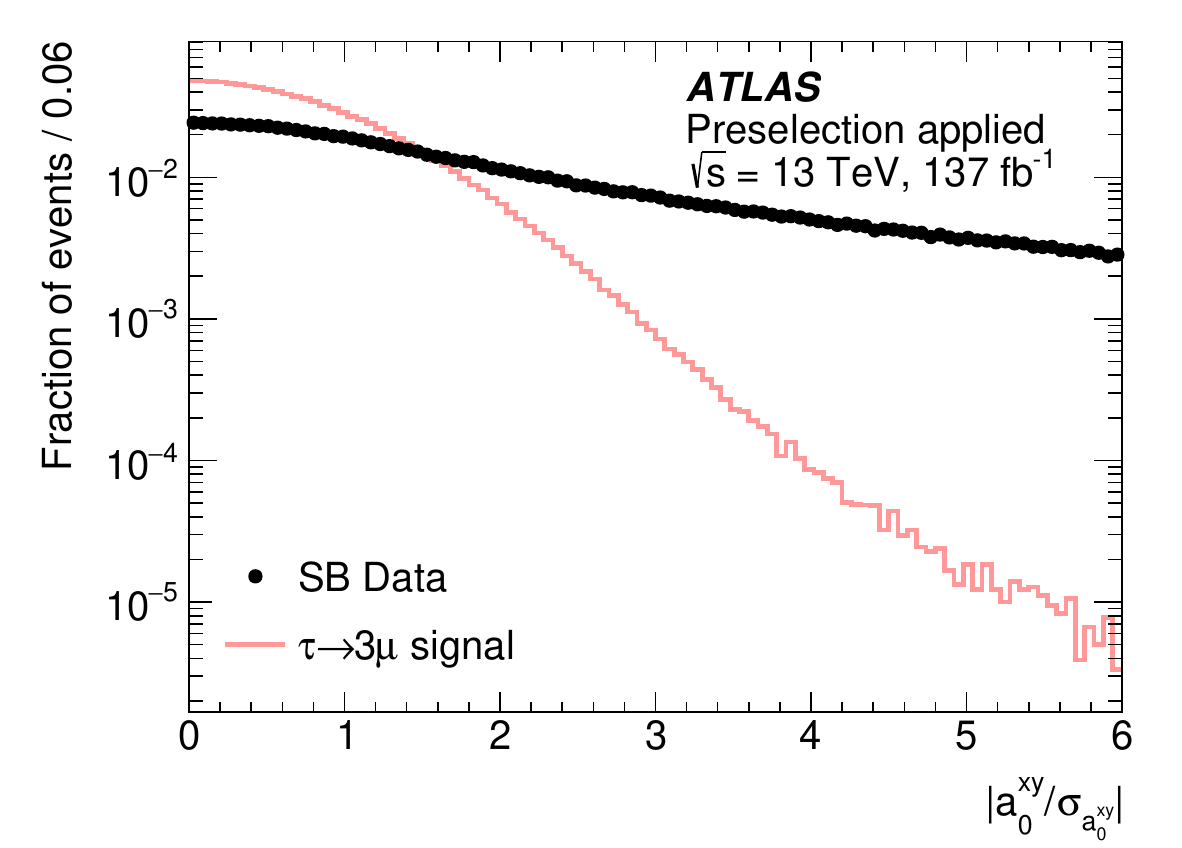}
}
\subfloat[]{
\includegraphics[width=0.45\linewidth]{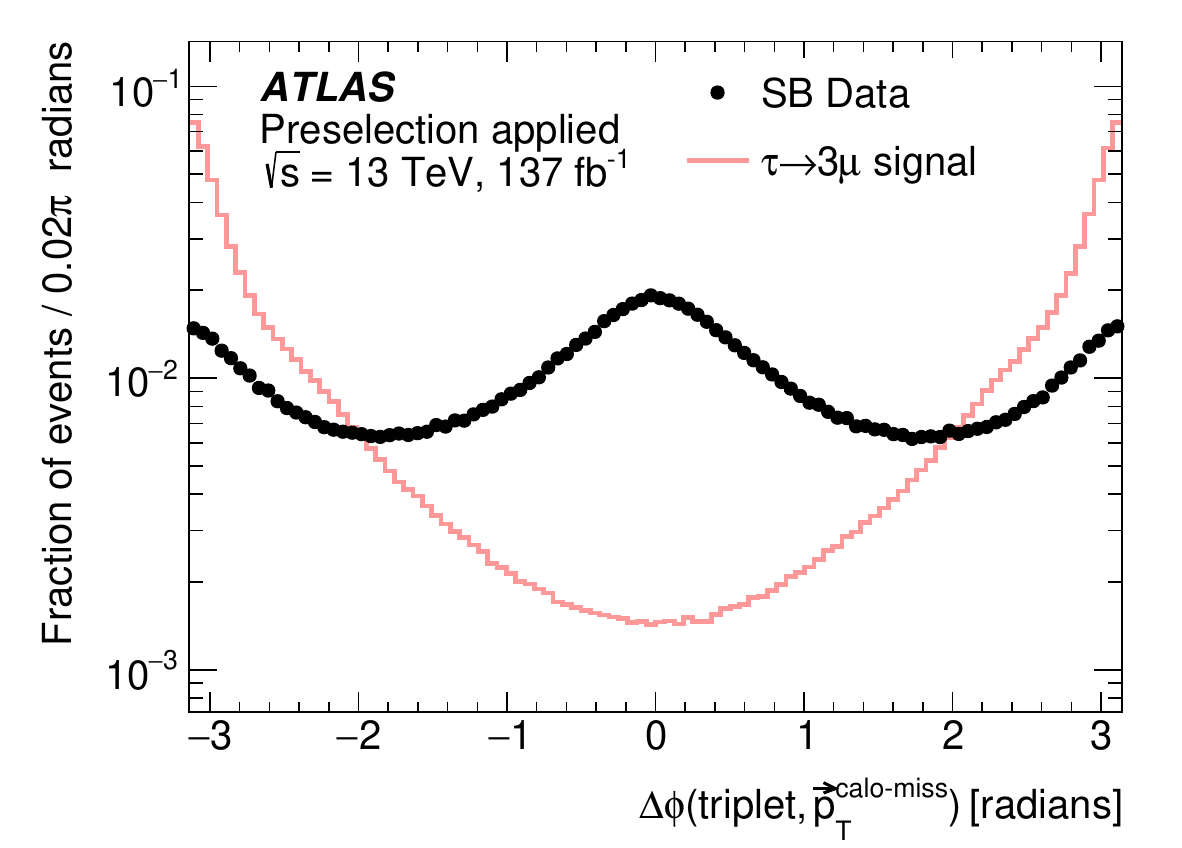}
}
\caption{Distributions of example variables used in the BDT event classification: (a) vertex quality $p\text{-value}$, (b) muon isolation $\pt^\mathrm{iso}/\pt^\mathrm{triplet}$, (c) triplet $\pt$, $\pt^{\text{triplet}}$, (d) significance of the transverse distance between the PV and SV $L^{\text{xy}}/\sigma_{L^{\text{xy}}}$, (e) absolute value of the significance of the transverse SV IP $|a_0^{xy}/\sigma_{a_0^{xy}}|$ and (f) azimuthal angle between the triplet and the calorimeter-based \met $\Delta\phi(\text{triplet}, \vec{p}_\mathrm{T}^{\text{calo-miss}})$. The plots show both the EW $\tau \to 3 \mu$ signal sample (histogram) and data from the sidebands (SB Data) (markers) after the preselection cuts. All distributions are normalised to unit area.}
\label{bdtplots}
\end{figure}

The resulting BDT score distributions for simulated signal events and
background events from the data sidebands are shown in
Figure~\ref{bdtdist}.
Good discrimination between signal and
background events is observed; additionally no significant over-training was found.

\begin{figure}
\centering
\includegraphics[width=0.65\linewidth]{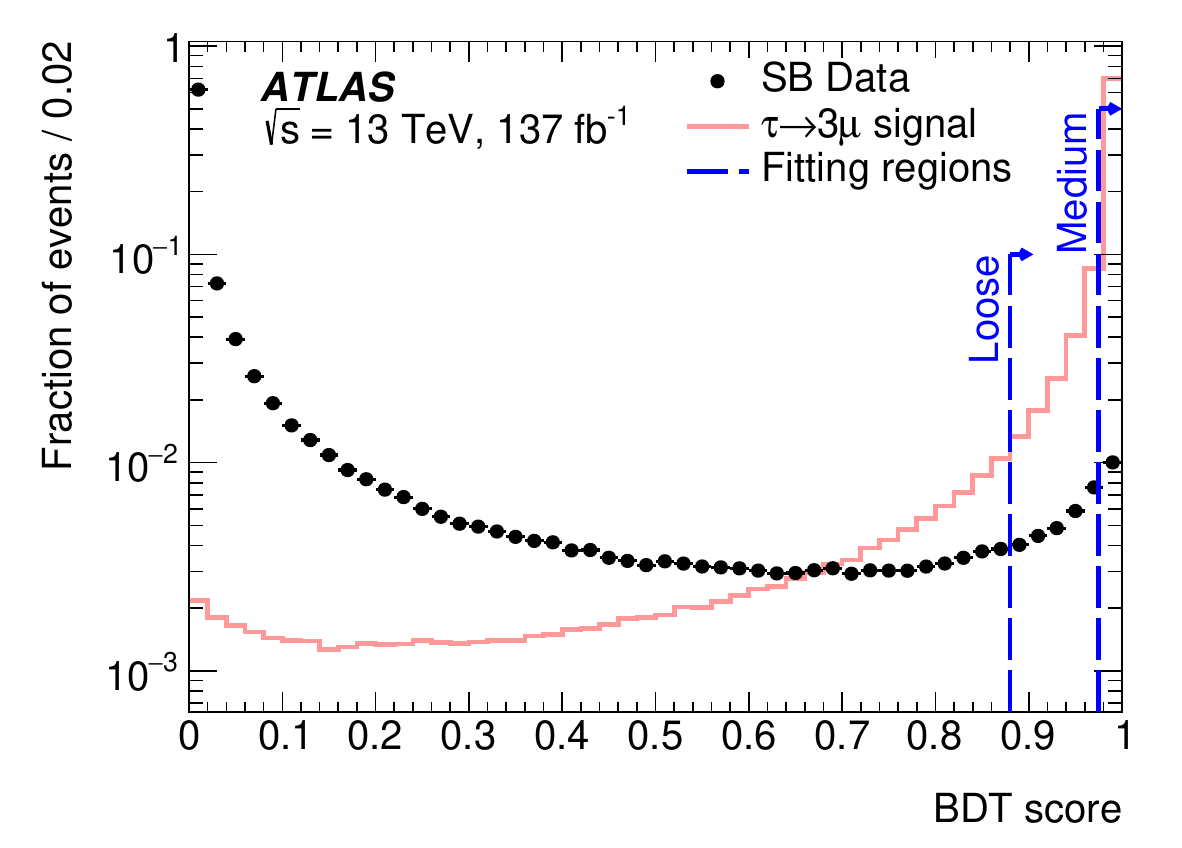}
\caption{BDT score distribution for the EW $\tau \to 3 \mu$
signal sample (histogram), and background from sideband data (SB Data) (markers).
All distributions are normalised to
unit area, and the error bars are too small to be visible. The vertical lines mark the boundaries of the
loose ($0.88<$BDT$\leq0.975$) and medium ($0.975<$BDT$\leq0.998$) BDT bins used for the event
categorisation. The tight boundary ($0.998<$BDT) is not shown as it is indistinguishable from the vertical axis.}
\label{bdtdist}
\end{figure}

Events are split into different categories to optimise the
sensitivity based on the BDT output score and different $m_{3\mu}$
detector resolutions. Events in the barrel, defined as
$|\eta_{\text{triplet}}|<1.2$, are found to have a signal mass
resolution of approximately $25\,\MeV$, significantly better than the
resolution, about $34\,\MeV$, for the remainder of the events that are assigned to the endcap category. Therefore, barrel and endcap events are separated.
To find the optimal BDT categories, the expected
limit on $B(\tau \to 3 \mu)$ is determined for a range of different BDT
categories by performing a simultaneous unbinned maximum-likelihood
fit on the $m_{3\mu}$ distribution in all
fit categories. The optimal limit is achieved using three BDT categories,
referred to as loose, medium and tight, as indicated on
Figure~\ref{bdtdist}. Table~\ref{tab:yields} shows the expected and observed signal and background yields in all six categories. In the most-sensitive tight categories, the $W \rightarrow
\tau{\nu}_\tau$ process contributes approximately $85\%$ of the EW signal.

\begin{table}[tb]
\caption{Number of events in each fit category. The top two lines (EW and HF signal) correspond to the expected signal for $B(\tau \to 3 \mu)  = 1\times 10^{-7}$, while in the lower two lines the fitted background contributions and the observed data events in the signal region ($1700\,\MeV < m_{3\mu} < 1850\,\MeV$) are listed.} \label{tab:yields}
\centering
\begin{tabular}{lrrrrrr}
\hline
& \multicolumn{6}{c}{Events in fit category}                          \\
\hline
\multicolumn{1}{r}{BDT category}    & \multicolumn{2}{c}{Loose} & \multicolumn{2}{c}{Medium} & \multicolumn{2}{c}{Tight} \\
Source & Barrel & Endcap & Barrel & Endcap & Barrel & Endcap \\
\hline
EW signal        & 1.0            &  2.0          & 3.2             & 5.8            & 2.5            & 5.3           \\
HF signal        & 1.0           & 2.4           & 0.6             & 1.4             & 0.1            & 0.2            \\ \hline
Background in signal region & 191.9 & 275.8 & 88.0  & 135.6  & 3.4 & 6.4   \\
Signal region data       & 209            & 291            & 86             & 141             & 5            & 5     \\
\hline
\end{tabular}
\end{table}

The simulated signals are parameterised separately in each category
with a double sided Crystal Ball (DSCB)
function~\autocite{crystalball,HIGG-2014-04}, which models the signal
$m_{3\mu}$ distribution well. The background $m_{3\mu}$ shape is
parameterised by a smooth function, chosen as an exponential, with the exponent
allowed to be positive or negative. This is
determined to be a suitable model in two ways. First, it is tested if the
background model is able to describe the data sidebands. Second,
signal-plus-background fits are performed to signal-depleted regions
of low BDT score, defined as the point below which less than 10\% of
the signal would be retained. The measured signal yield in this
region, known as the spurious signal~\autocite{HIGG-2013-08},
is found to be suitable following the criteria in Ref.\autocite{ATL-PHYS-PUB-2020-028} for the chosen functional form.


\section{Systematic uncertainties} \label{sec:systematics}

The analysis is found to be limited by the statistical uncertainty of the dataset.
Experimental systematic uncertainties apply to the
simulated signal, as the background is determined directly from the
data. Further cross-checks on the impact of the choice of background
parameterisation are reported at the end of this Section, including a
dedicated systematic uncertainty.

The main source of systematic uncertainty is from the trigger efficiency,
determined as described in Section~\ref{sec:selection}.
Additional uncertainties are also considered for: the muon reconstruction
efficiency as determined in $J/\psi \rightarrow \mu \mu$ and
$Z \rightarrow \mu \mu$ events using a tag-and-probe
method~\autocite{MUON-2018-03}; the jet energy
scale~\autocite{JETM-2018-05}; the resolution of the
\met~\autocite{JETM-2020-03}; and the level of
pile-up~\autocite{STDM-2015-05}. These effects vary in size between
the fit categories and correlations are considered. The effect on the
signal normalisation for each category is shown in
Table~\ref{systematictable}.  Other systematic variations are found to
have an effect of less than $1\%$ and are neglected.

Further sources of systematic uncertainties have a global
effect on the normalisation. The uncertainty in the integrated
luminosity is 0.84\%~\autocite{DAPR-2021-01},
obtained using the LUCID-2 detector \autocite{LUCID2} for the primary
luminosity measurements, complemented by measurements using the inner
detector and calorimeters. The uncertainty on the EW signal cross-section is
conservatively taken as the maximum fractional uncertainty in the $W \to \ell \nu$,
$Z \to \ell \ell$ and $t\bar{t}$ cross-section
measurements~\autocite{STDM-2018-17, TOPQ-2018-26}, which is $3.0\%$. For the
HF signal, the uncertainty in the $D_s$ inclusive production
cross-section \autocite{BPHY-2018-10} of $14\%$ is used along with a
$1.9\%$ uncertainty in the $B(D_s \to \tau\nu_\tau)$
\autocite{Zyla:2020zbs} and a $12\%$ uncertainty in the
$B(B \to \tau X)$~\autocite{CMS-BPH-17-004}. The impact of
further simulation modelling uncertainties are expected to be
negligible and are not considered.

\begin{table}[]
\caption{Signal normalisation uncertainties from different
systematic variations for the six fit categories. Only effects
with a magnitude of at least $1\%$ in at least one category are
reported. The effect of each systematic variation is treated as correlated between categories. Systematic effects shown with different signs refer to migration effects that lead to anti-correlations between the yields of different categories. Further global
signal normalisation effects are discussed in the text.}
\centering
\begin{tabular}{lrrrrrr}
\hline
& \multicolumn{6}{c}{Systematic signal uncertainty [\%]}                          \\
\hline
\multicolumn{1}{r}{BDT category}    & \multicolumn{2}{c}{Loose} & \multicolumn{2}{c}{Medium} & \multicolumn{2}{c}{Tight} \\
Systematic uncertainty & \multicolumn{1}{c}{Barrel} & \multicolumn{1}{c}{Endcap} & \multicolumn{1}{c}{Barrel} & \multicolumn{1}{c}{Endcap} & \multicolumn{1}{c}{Barrel} & \multicolumn{1}{c}{Endcap}\\
\hline
Trigger efficiency    & $\pm 26.7$        & $\pm 10.1$        & $\pm 20.9$         & $\pm 9.1$          & $\pm 16.2$        & $\pm 11.4$        \\
Muon reconstruction efficiency & $\pm 3.7$ & $\pm 3.5$ & $\pm 3.4$ & $\pm 3.2$ & $\pm 2.9$ & $\pm 3.0$ \\
\met track resolution & $\pm 1.9$ & $\pm 1.1$ & $\pm 0.8$ & $\pm 1.2$ & $\mp 2.4$ & $\mp 2.3$ \\
Jet energy scale & $\pm 5.0$ & $\pm 3.3$ & $\pm 0.9$ & $\pm 0.4$ & $\mp 5.1$ & $\mp 3.2$ \\
Pile-up     & $\pm 1.5$         & $\pm 1.5$         & $<0.1$         & $<0.1$          & $\mp 1.8$        & $\mp 1.4$       \\
\hline
\end{tabular}
\label{systematictable}
\end{table}

The effect of the systematic uncertainties in the signal shape is also
considered, with the muon momentum calibration~\autocite{MUON-2022-01}
being the only variation having an effect.
The muon momentum scale uncertainty has a maximum effect on the mean of the signal
DSCB of $0.1\%$ in the barrel and $0.3\%$ in the endcap. The
momentum resolution has a maximum effect on the width of the signal
DSCB of $3.2\%$ ($0.8\%$) in the barrel and $1.6\%$ ($3.3\%$) in the endcap for the ID (MS) contribution.

The impact of a different choice of background parameterisation on the
number of background events has been investigated by replacing the nominal
exponential function by a second-order polynomial or a power law. The
different functions resulted in a maximum $5\%$ change in the number of extracted
background events, well within the statistical
uncertainty on the background extraction. Constraining
the background to be constant or falling with increasing $m_{3\mu}$
results in no systematic change in the number of background events.
Although the magnitude of the spurious signal uncertainty is found to be suitable following the criteria in
Ref.~\autocite{ATL-PHYS-PUB-2020-028} for the chosen functional form, a systematic uncertainty to
account for possible residual mismodelling effects is applied. The spurious signal yield,
in the signal-depleted region mentioned previously, is scaled for each
of the fit categories and treated as an uncorrelated systematic uncertainty. This background parameterisation
uncertainty only has a small impact on the final results.


%

\section{Results} \label{sec:result}

To find evidence for the decay of $\tau \rightarrow 3 \mu$ or to set a
limit on the branching ratio, a simultaneous unbinned maximum-likelihood
fit is performed on the $m_{3\mu}$ distribution in all six
fit categories. The sum of EW and HF signals are parameterised in
each category separately with a DSCB, while the background is parameterised with an
exponential function. The systematic uncertainties as discussed in Section~\ref{sec:systematics} are implemented as
nuisance parameters in the fit. For these fits, $B(\tau \rightarrow 3 \mu)$ is
varied as the parameter of interest, and the profile likelihood test statistic in the asymptotic
approximation \autocite{Cowan:2010js} is used. Injection tests with signals in the
range of $B(\tau \rightarrow 3 \mu) = 10^{-8}$--$10^{-6}$ are performed to verify that the fit is unbiased.

\begin{figure}
\centering
\subfloat[]{
\includegraphics[width=0.45\linewidth]{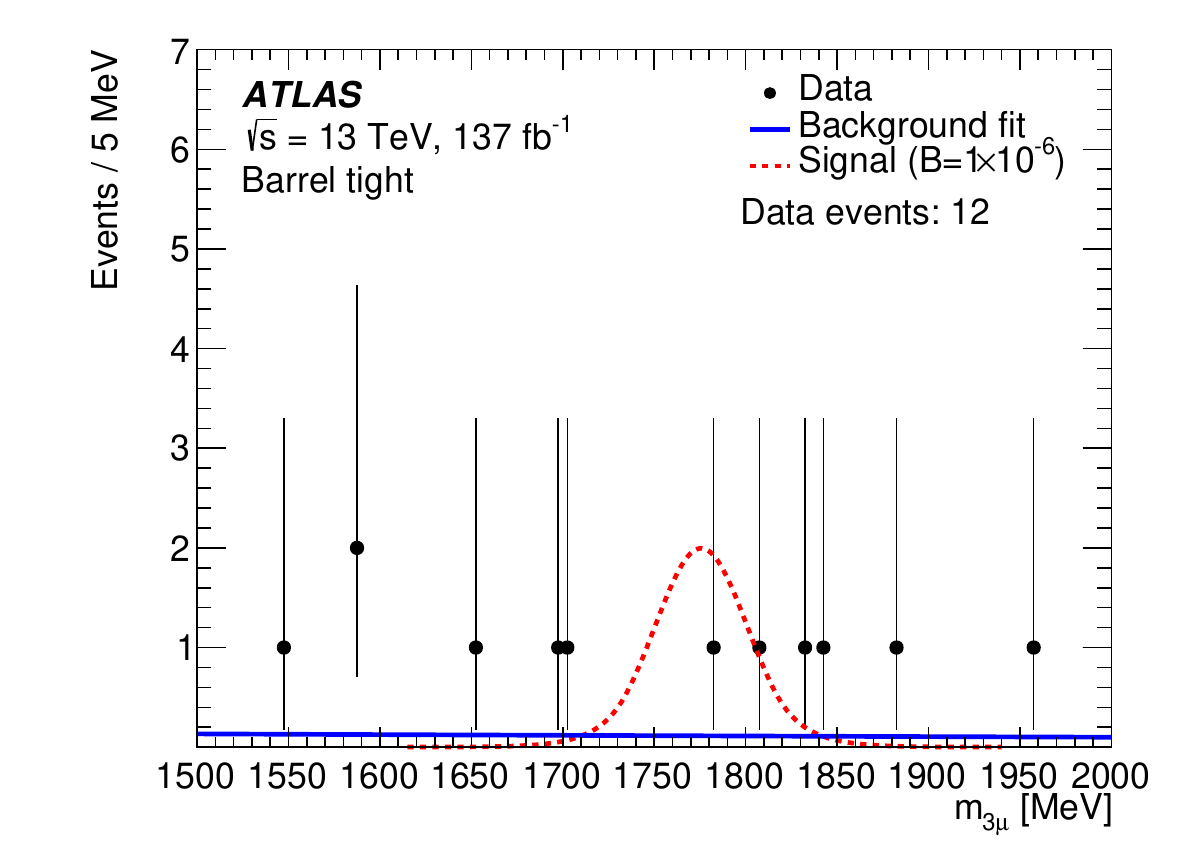}
}
\subfloat[]{
\includegraphics[width=0.45\linewidth]{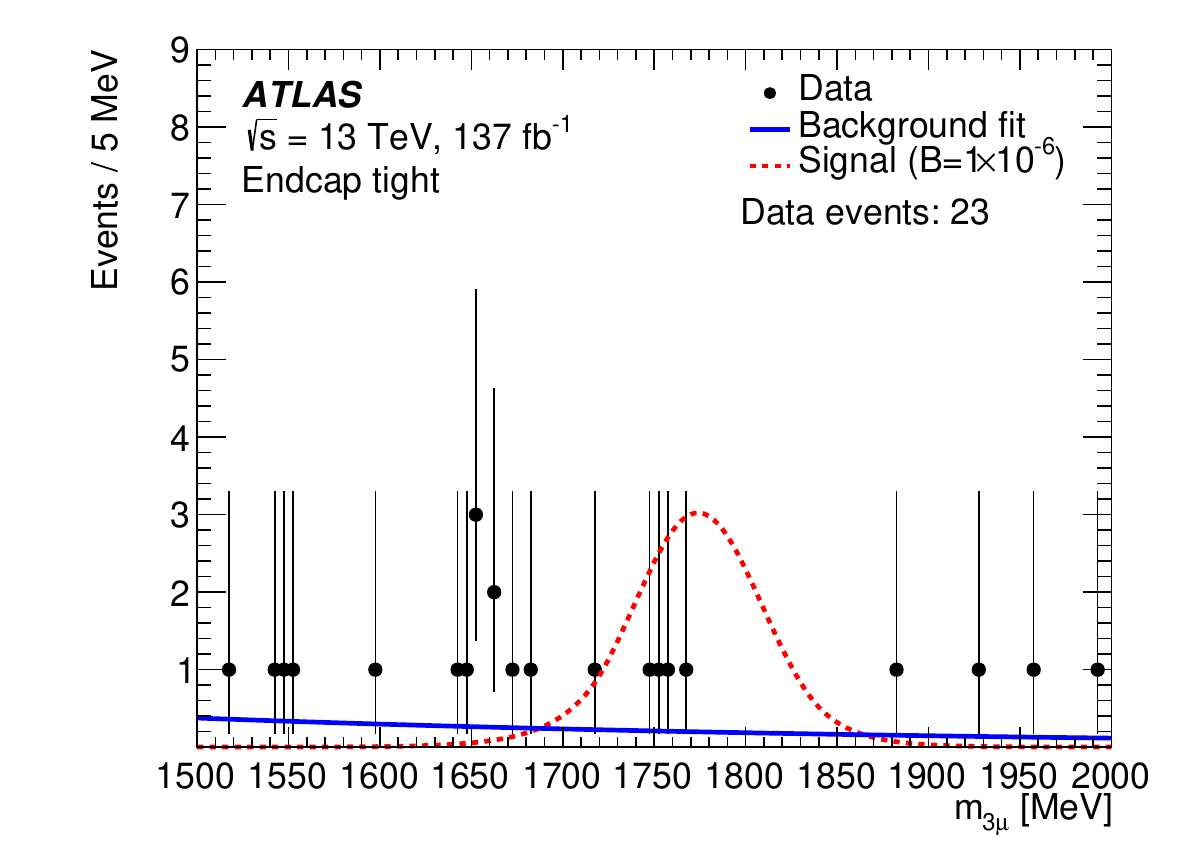}
} \hfill
\subfloat[]{
\includegraphics[width=0.45\linewidth]{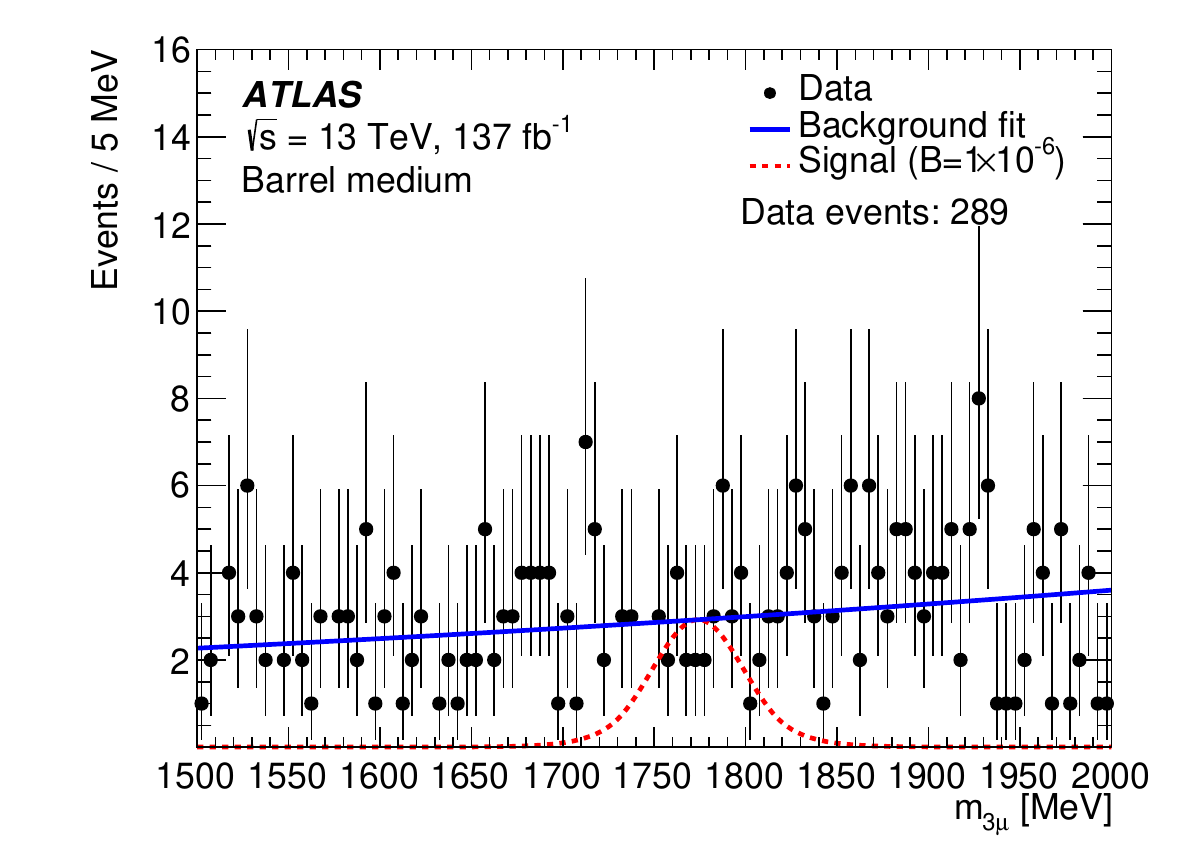}
}
\subfloat[]{
\includegraphics[width=0.45\linewidth]{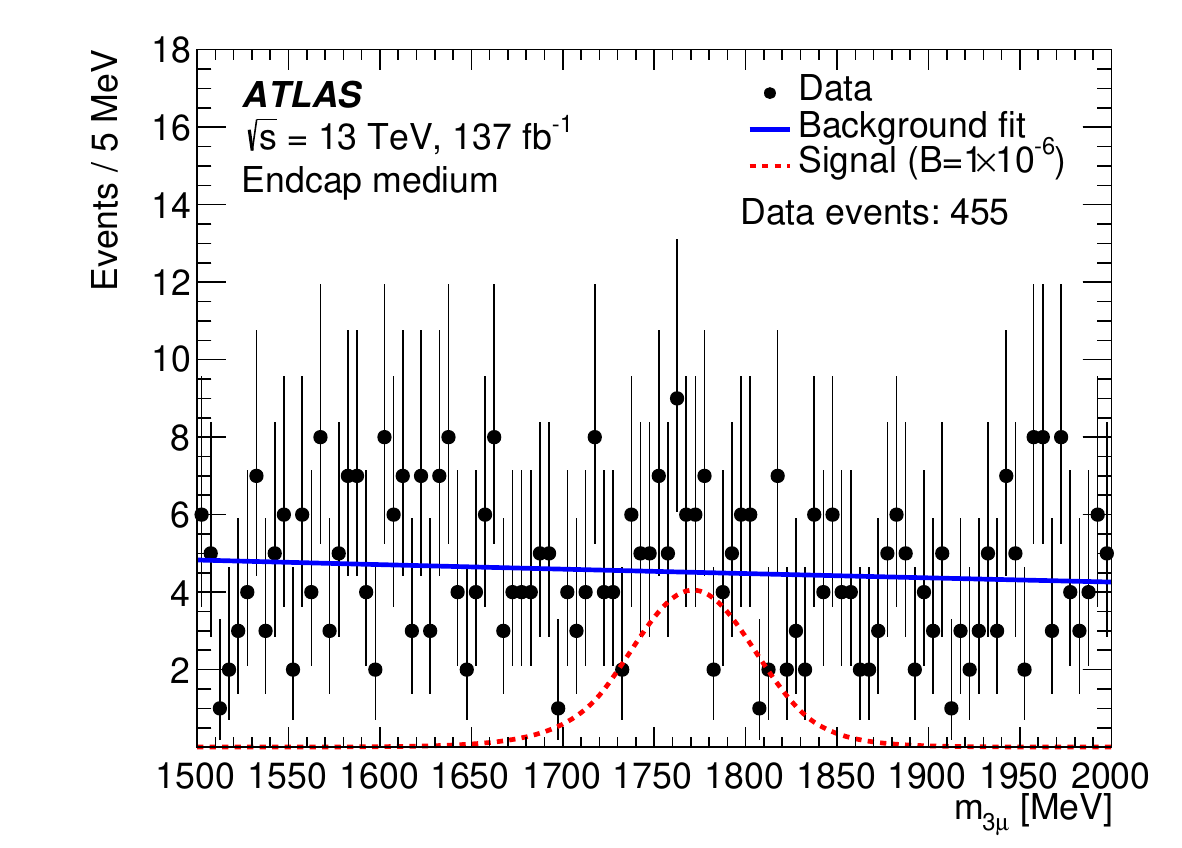}
} \hfill
\subfloat[]{
\includegraphics[width=0.45\linewidth]{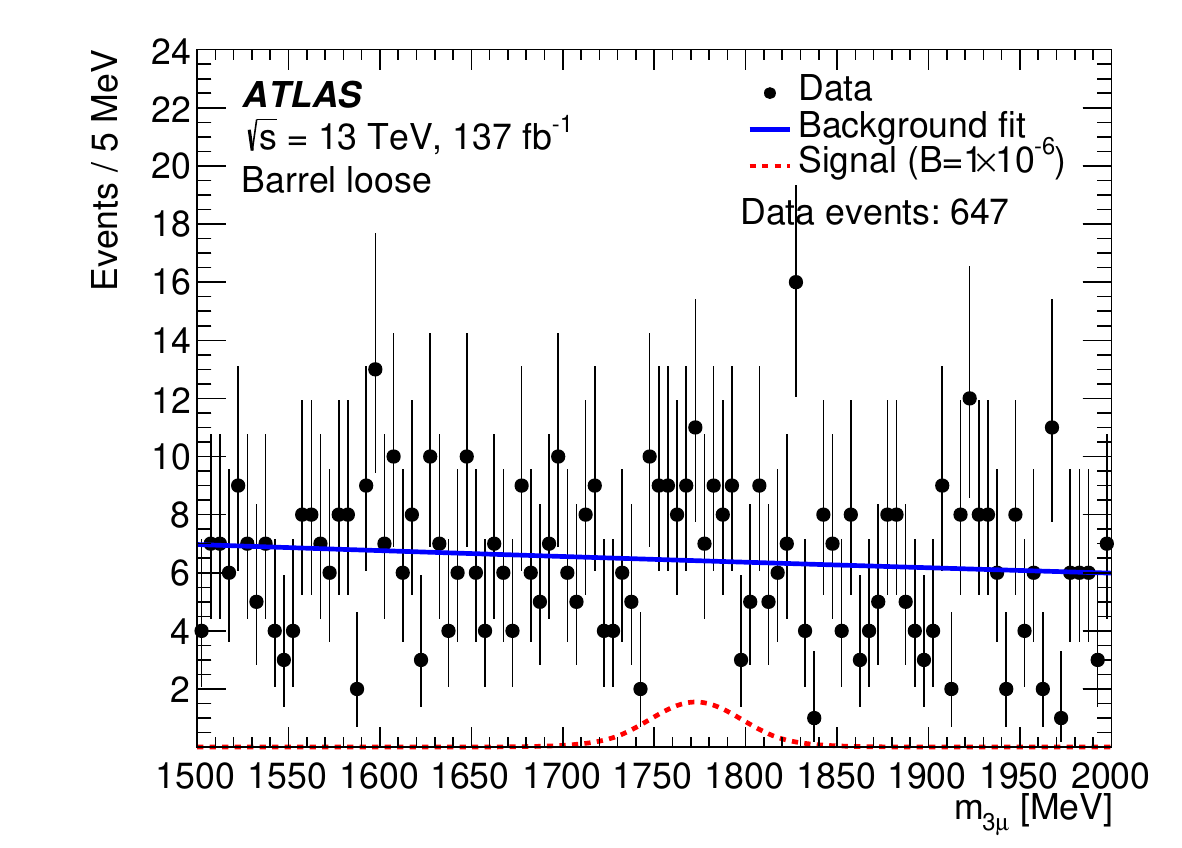}
}
\subfloat[]{
\includegraphics[width=0.45\linewidth]{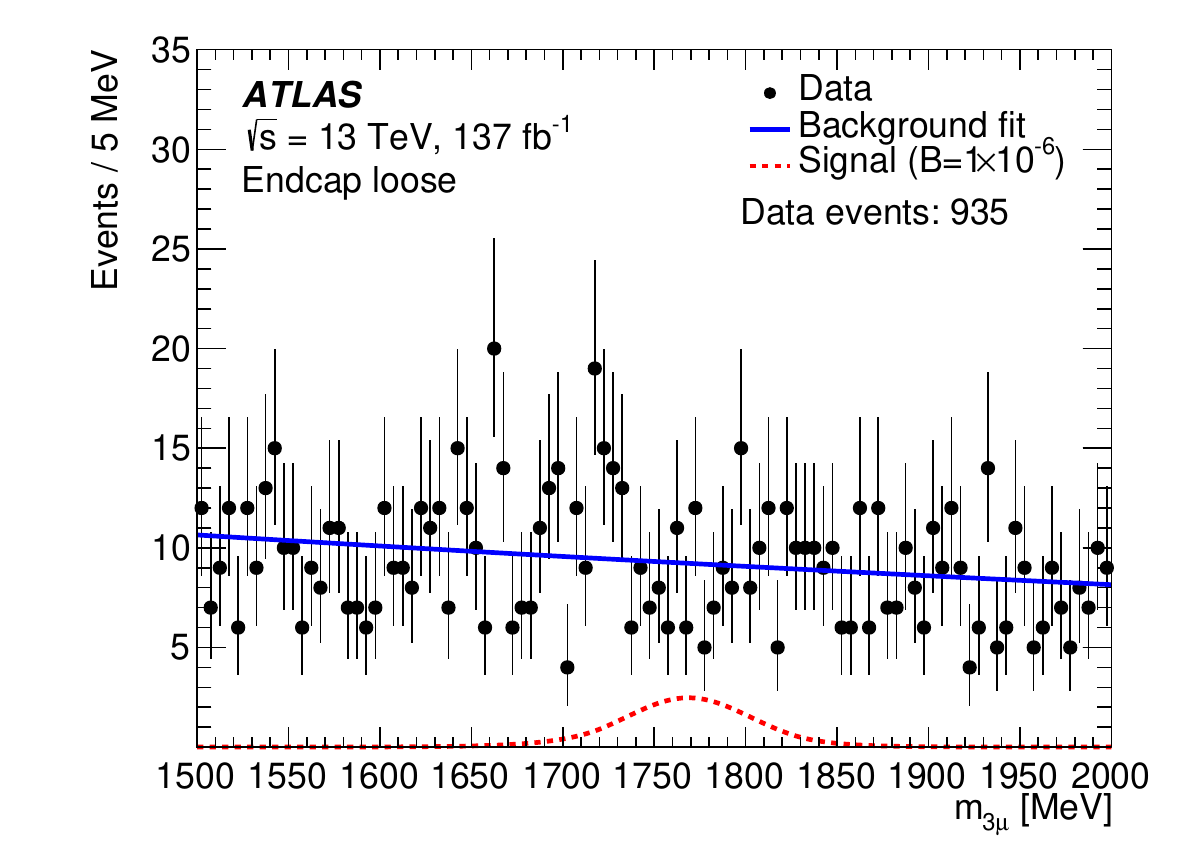}
}

\caption{The $m_{3\mu}$ in data (markers), background-only fit (solid line) and
expected EW plus HF signal, for $B(\tau \to 3 \mu)  = 1\times 10^{-6}$ as an illustration (dotted line) in the (a,c,e) barrel and (b,d,f) endcap categories
for the (a,b) tight, (c,d) medium and (e,f) loose BDT
categories.}
\label{unblindfitting}
\end{figure}

Figure~\ref{unblindfitting} shows the background-only fit and the expected
signal, for $B(\tau \to 3 \mu)  = 1\times 10^{-6}$, in the unblinded $m_{3\mu}$ spectra. No significant signal is
observed and the best-fit signal strength value is $B(\tau \to 3\mu) = (1.5 \pm 4.0) \times 10^{-8}$.
An upper limit is set using the CL$_\text{s}$ method
\autocite{Read:2002hq}. The observed (expected) upper limit is found
to be $B(\tau \to 3\mu) < \limcombobsstandard \times 10^{-8}$
($\limcombexpstandard \times 10^{-8}$) at $90\%$~CL, with the HF signal contributing $6\%$
to the expected limit. The
upper limit is also calculated at $95\%$ CL, resulting in an observed
(expected) upper limit of $\limcombobsCarl \times 10^{-8}$
($\limcombexpCarl \times 10^{-8}$). The CL$_\text{s}$ limit scan is
shown in Figure~\ref{limitscan}, with the observed limit well within $1$~$\sigma$ of the expected limit. %

Using pseudo experiments to calculate the limit instead of the asymptotic approximation is found to have
a negligible impact on the limits obtained: the observed limit improves by $1\%$, while the
expected limit worsens by $3\%$. Excluding all systematic uncertainties results
in an improvement to the expected limit by $4.4\%$, with the trigger
efficiency systematic uncertainty having the largest contribution of $2.1\%$,
followed by the spurious signal systematic uncertainty contribution of $1.8\%$.

\begin{figure}
\centering
\includegraphics[width=0.75\linewidth]{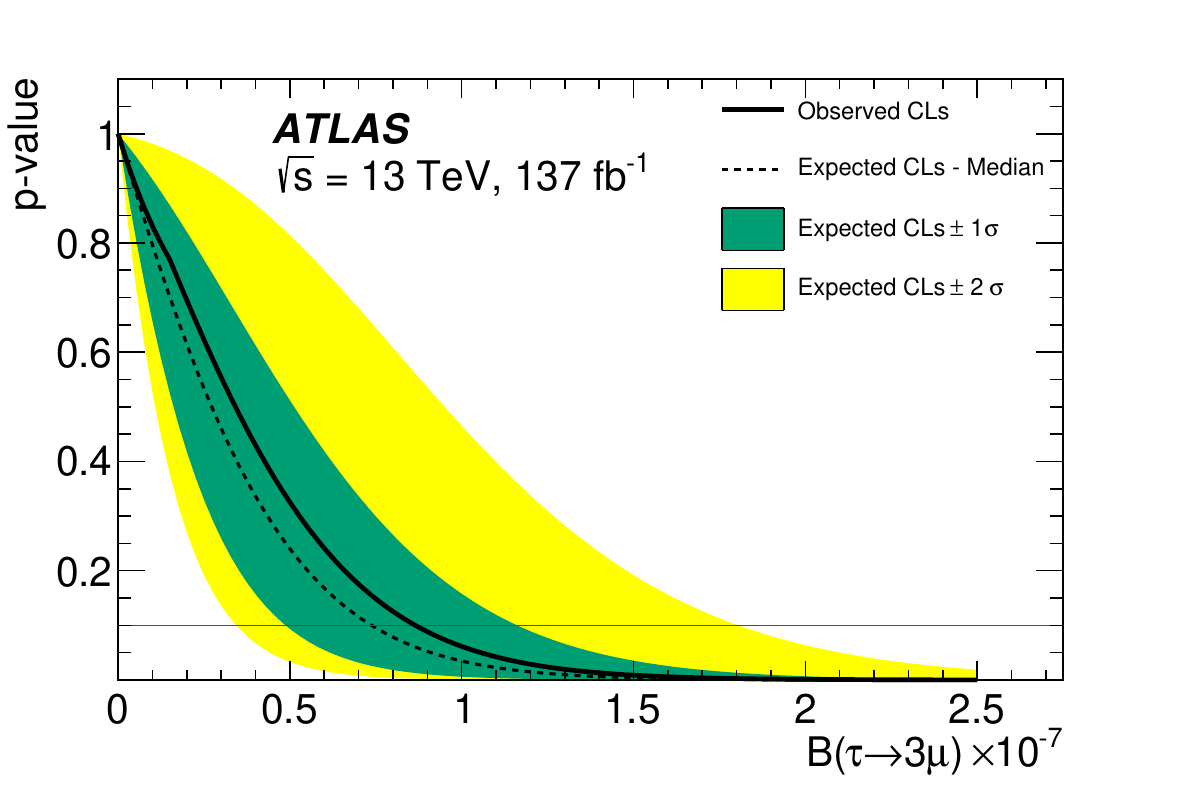}
\caption{$B(\tau\to3\mu)$ limit scan showing the expected limit (dotted line), $\pm 1(2) \sigma$ error bands (green and yellow), and the observed limit using the CL$_\text{s}$ method (solid line). A
$p$-value of $0.1$ ($90\%$ CL) is shown by the horizontal line.}
\label{limitscan}
\end{figure}


%
%
\FloatBarrier

%

\section{Conclusion}

This paper presents a search for the cLFV process $\tau \rightarrow
3\mu$, using $137\,\ifb$ of $pp$ collision data collected by the ATLAS
experiment during Run~2 of the Large Hadron Collider at $\sqrt{s} =
13\,\TeV$. The analysis is optimised using MVA analysis techniques
for electroweak sources, but also considers  $\tau$-lepton production from
heavy-flavour sources. No evidence for $\tau \rightarrow
3\mu$ decays is found and observed (expected) limits are set on
$B(\tau \rightarrow 3\mu) < \limcombobsstandard\times 10^{-8}$ $(\limcombexpstandard \times 10^{-8})$ at $90\%$~CL. This result
improves upon the previous ATLAS Run 1 limit by a factor of 5, where a
factor of 3 is expected solely from the larger production cross-sections
at the increased centre-of-mass energy and larger integrated
luminosity. The additional improvement is attributed to detector
improvements such as the new IBL detector, improved machine learning
methods, and new more efficient triggers with lower \pt thresholds. The analysis is limited by statistical uncertainties
and thus improvements in sensitivity are expected with the analysis
of additional data.


%
%
%
%
%
\section*{Acknowledgements}
%

%

%
%

%
%

We thank CERN for the very successful operation of the LHC and its injectors, as well as the support staff at
CERN and at our institutions worldwide without whom ATLAS could not be operated efficiently.

The crucial computing support from all WLCG partners is acknowledged gratefully, in particular from CERN, the ATLAS Tier-1 facilities at TRIUMF/SFU (Canada), NDGF (Denmark, Norway, Sweden), CC-IN2P3 (France), KIT/GridKA (Germany), INFN-CNAF (Italy), NL-T1 (Netherlands), PIC (Spain), RAL (UK) and BNL (USA), the Tier-2 facilities worldwide and large non-WLCG resource providers. Major contributors of computing resources are listed in Ref.~\cite{ATL-SOFT-PUB-2026-001}.

We gratefully acknowledge the support of ANPCyT, Argentina; YerPhI, Armenia; ARC, Australia; BMWFW and FWF, Austria; ANAS, Azerbaijan; CNPq and FAPESP, Brazil; NSERC, NRC and CFI, Canada; CERN; ANID, Chile; CAS, MOST and NSFC, China; Minciencias, Colombia; MEYS CR, Czech Republic; DNRF and DNSRC, Denmark; IN2P3-CNRS and CEA-DRF/IRFU, France; SRNSFG, Georgia; BMFTR, HGF and MPG, Germany; GSRI, Greece; RGC and Hong Kong SAR, China; ICHEP and Academy of Sciences and Humanities, Israel; INFN, Italy; MEXT and JSPS, Japan; CNRST, Morocco; NWO, Netherlands; RCN, Norway; MNiSW, Poland; FCT, Portugal; MNE/IFA, Romania; MSTDI, Serbia; MSSR, Slovakia; ARIS and MVZI, Slovenia; DSI/NRF, South Africa; MICIU/AEI, Spain; SRC and Wallenberg Foundation, Sweden; SERI, SNSF and Cantons of Bern and Geneva, Switzerland; NSTC, Taipei; TENMAK, T\"urkiye; STFC/UKRI, United Kingdom; DOE and NSF, United States of America.

Individual groups and members have received support from BCKDF, CANARIE, CRC and DRAC, Canada; CERN-CZ, FORTE and PRIMUS, Czech Republic; COST, ERC, ERDF, Horizon 2020 and Marie Sk{\l}odowska-Curie Actions, European Union; Investissements d'Avenir Labex, Investissements d'Avenir Idex and ANR, France; DFG and AvH Foundation, Germany; Herakleitos, Thales and Aristeia programmes co-financed by EU-ESF and the Greek NSRF, Greece; BSF-NSF and MINERVA, Israel; NCN and NAWA, Poland; La Caixa Banking Foundation, CERCA and AGAUR programs from Generalitat de Catalunya and PROMETEO and GenT Programmes Generalitat Valenciana, Spain; G\"{o}ran Gustafssons Stiftelse, Sweden; The Royal Society and Leverhulme Trust, United Kingdom; Eric and Wendy Schmidt Fund for Strategic Innovation, United States of America.

In addition, individual members wish to acknowledge support from Chile: Agencia Nacional de Investigaci\'on y Desarrollo (ANID FONDECYT reg. 1230987, FONDECYT 1230812, FONDECYT 1240864, Fondecyt 3240661, Fondecyt Regular 1240721); China: Chinese Ministry of Science and Technology (MOST-2023YFA1605700, MOST-2023YFA1609300), National Natural Science Foundation of China (NSFC 12275265, NSFC-W2543005); Czech Republic: Czech Science Foundation (GACR - 24-11373S), Ministry of Education Youth and Sports (ERC-CZ-LL2327, FORTE CZ.02.01.01/00/22\_008/0004632), PRIMUS Research Programme (PRIMUS/21/SCI/017); EU: H2020 European Research Council (ERC - 101002463); European Union: European Research Council (BARD No. 101116429, ERC - 101219398, ERC - 948254, ERC 101089007), European Regional Development Fund (HE COFUND GA No.101081355, ERDF), Marie Sklodowska-Curie Actions (GAP-101168829); France: Agence Nationale de la Recherche (ANR-21-CE31-0013, ANR-22-EDIR-0002, ANR-24-CE31-0504-01); Germany: Deutsche Forschungsgemeinschaft (DFG - 469666862); China: Research Grants Council (GRF); Italy: Istituto Nazionale di Fisica Nucleare (LHC-MIUR - 28003/2025), Ministero dell'Università e della Ricerca (NextGenEU 153D23001490006 M4C2.1.1, NextGenEU I53D23000820006 M4C2.1.1, NextGenEU I53D23001490006 M4C2.1.1, SOE2024\_0000023); Japan: Japan Society for the Promotion of Science (JSPS KAKENHI  JP25H0063, JSPS KAKENHI JP22H01227, JSPS KAKENHI JP22H04944, JSPS KAKENHI JP22KK0227, JSPS KAKENHI JP24K23939, JSPS KAKENHI JP24KK0251, JSPS KAKENHI JP25H00650, JSPS KAKENHI JP25H01291, JSPS KAKENHI JP25K01011, JSPS KAKENHI JP25K01023); Norway: Research Council of Norway (RCN-314472); Poland: Polish National Science Centre (NCN 2021/42/E/ST2/00350, NCN OPUS 2023/51/B/ST2/02507, NCN OPUS nr 2022/47/B/ST2/03059, NCN UMO-2019/34/E/ST2/00393, UMO-2022/47/O/ST2/00148, UMO-2023/49/B/ST2/04085, UMO-2023/51/B/ST2/00920, UMO-2024/53/N/ST2/00869); Spain: Agència de Gestió d'Ajuts Universitaris i de Recerca. (AGAUR - 2023 BP 00141), Ministry of Science and Innovation (RYC2019-028510-I, RYC2020-030254-I, RYC2021-031273-I, RYC2022-038164-I), Ministerio de Ciencia, Innovación y Universidades/Agencia Estatal de Investigaci\'on (EU NextGenerationEU (PRTR-C17.I1), PID2022-142604OB-C22); Sweden: Carl Trygger Foundation (Carl Trygger Foundation CTS 22:2312), Swedish Research Council (Swedish Research Council 2023-04654, VR 2021-03651, VR 2022-03845, VR 2022-04683, VR 2023-03403, VR 2024-05451, VR 2025-05940), Knut and Alice Wallenberg Foundation (KAW 2023.0366); Switzerland: Swiss National Science Foundation (SNSF - PCEFP2\_194658); United Kingdom: The Binks Trust, Royal Society (NIF-R1-231091); United States of America: U.S. Department of Energy (ECA DE-AC02-76SF00515), John Templeton Foundation (John Templeton Foundation 63206), Neubauer Family Foundation.

%
%


%
\clearpage
\printbibliography
\clearpage
\input{atlas_authlist}
%

%
%

%

%
%
%
%

\end{document}

%% file: atlas_authlist.tex
 
\begin{flushleft}
\hypersetup{urlcolor=black}
{\Large The ATLAS Collaboration}

\bigskip

\AtlasOrcid[0000-0002-6665-4934]{G.~Aad}$^\textrm{\scriptsize 102}$,
\AtlasOrcid[0000-0001-7616-1554]{E.~Aakvaag}$^\textrm{\scriptsize 17}$,
\AtlasOrcid[0000-0002-5888-2734]{B.~Abbott}$^\textrm{\scriptsize 121}$,
\AtlasOrcid[0000-0002-0287-5869]{S.~Abdelhameed}$^\textrm{\scriptsize 83b}$,
\AtlasOrcid[0000-0002-1002-1652]{K.~Abeling}$^\textrm{\scriptsize 54}$,
\AtlasOrcid[0000-0001-5763-2760]{N.J.~Abicht}$^\textrm{\scriptsize 48}$,
\AtlasOrcid[0000-0002-8496-9294]{S.H.~Abidi}$^\textrm{\scriptsize 30}$,
\AtlasOrcid[0009-0003-6578-220X]{M.~Aboelela}$^\textrm{\scriptsize 44}$,
\AtlasOrcid[0000-0002-9987-2292]{A.~Aboulhorma}$^\textrm{\scriptsize 36e}$,
\AtlasOrcid[0000-0001-5329-6640]{H.~Abramowicz}$^\textrm{\scriptsize 154}$
% ...
\bigskip
\\

% ...
$^{17}$Department for Physics and Technology, University of Bergen, Bergen; Norway.\\
% ...
$^{30}$Physics Department, Brookhaven National Laboratory, Upton NY; United States of America.\\
% ...
$^{36}$$^{(a)}$Facult\'e des Sciences Ain Chock, Universit\'e Hassan II de Casablanca;$^{(b)}$Facult\'{e} des Sciences, Universit\'{e} Ibn-Tofail, K\'{e}nitra;$^{(c)}$Facult\'e des Sciences Semlalia, Universit\'e Cadi Ayyad, LPHEA-Marrakech;$^{(d)}$LPMR, Facult\'e des Sciences, Universit\'e Mohamed Premier, Oujda;$^{(e)}$Facult\'e des sciences, Universit\'e Mohammed V, Rabat;$^{(f)}$Institute of Applied Physics, Mohammed VI Polytechnic University, Ben Guerir; Morocco.\\
% ...
$^{44}$Physics Department, Southern Methodist University, Dallas TX; United States of America.\\
% ...
$^{48}$Fakult\"{a}t Physik, Technische Universit{\"a}t Dortmund, Dortmund; Germany.\\
% ...
$^{54}$II. Physikalisches Institut, Georg-August-Universit\"{a}t G\"ottingen, G\"ottingen; Germany.\\
% ...
$^{83}$$^{(a)}$Khalifa University of Science and Technology, Abu Dhabi;$^{(b)}$New York University Abu Dhabi, Abu Dhabi;$^{(c)}$United Arab Emirates University, Al Ain;$^{(d)}$University of Sharjah, Sharjah; United Arab Emirates.\\
% ...
$^{102}$CPPM, Aix-Marseille Universit\'e, CNRS/IN2P3, Marseille; France.\\
% ...
$^{121}$Homer L. Dodge Department of Physics and Astronomy, University of Oklahoma, Norman OK; United States of America.\\
% ...
$^{154}$Raymond and Beverly Sackler School of Physics and Astronomy, Tel Aviv University, Tel Aviv; Israel.\\
% ...
$^{a}$ Also at Affiliated with an institute formerly covered by a cooperation agreement with CERN.\\
$^{b}$ Also at An-Najah National University, Nablus; Palestine.\\
$^{c}$ Also at Borough of Manhattan Community College, City University of New York, New York NY; United States of America.\\
$^{d}$ Also at Center for Interdisciplinary Research and Innovation (CIRI-AUTH), Thessaloniki; Greece.\\
$^{e}$ Also at Centre of Physics of the Universities of Minho and Porto (CF-UM-UP); Portugal.\\
$^{f}$ Also at CERN, Geneva; Switzerland.\\
% ...

\end{flushleft}


%% file: ANA-BPHY-2018-05-PAPER.bib
@article{Belle_II,  
    author =  {{Belle II Collaboration}},
    title = "{Search for lepton-flavor-violating {\ensuremath{\tau}}$^{-}${\textrightarrow} {\ensuremath{\mu}}$^{-}${\ensuremath{\mu}}$^{+}${\ensuremath{\mu}}$^{-}$ decays at Belle II}",
    eprint = "2405.07386",
    archivePrefix = "arXiv",
    primaryClass = "hep-ex",
    reportNumber = "Belle II Preprint 2024-012 KEK Preprint 2024-6",
    doi = "10.1007/JHEP09(2024)062",
    journal = "JHEP",
    volume = "09",
    pages = "062",
    year = "2024"
}

@article{Belle,  
    author =  {{Belle Collaboration}},
    title = "{Search for lepton-flavor-violating {\ensuremath{\tau}} decays into three leptons with 719 million produced {\ensuremath{\tau}}$^{+}${\ensuremath{\tau}}$^{-}$ pairs}",
    eprint = "1001.3221",
    archivePrefix = "arXiv",
    primaryClass = "hep-ex",
    doi = "10.1016/j.physletb.2010.03.037",
    journal = "Phys. Lett. B",
    volume = "687",
    pages = "139--143",
    year = "2010"
}

@article{Babar,
    author = {{BaBar Collaboration}},
    title = "{Limits on {\ensuremath{\tau}} lepton-flavor violating decays into three charged leptons}",
    eprint = "1002.4550",
    archivePrefix = "arXiv",
    primaryClass = "hep-ex",
    reportNumber = "BABAR-PUB-09-033, SLAC-PUB-14009",
    doi = "10.1103/PhysRevD.81.111101",
    journal = "Phys. Rev. D",
    volume = "81",
    pages = "111101",
    year = "2010"
}

@article{decay_tau_3mu_sm,
    author = "Hern{\'a}ndez-Tom{\'e}, G. and L{\'o}pez Castro, G. and Roig, P.",
    title = "{Flavor violating leptonic decays of $\tau$ and $\mu$ leptons in the Standard Model with massive neutrinos}",
    eprint = "1807.06050",
    archivePrefix = "arXiv",
    primaryClass = "hep-ph",
    doi = "10.1140/epjc/s10052-019-6563-4",
    journal = "Eur. Phys. J. C",
    volume = "79",
    number = "1",
    pages = "84",
    year = "2019",
    related        = "decay_tau_3mu_sm-err",
    relatedstring  = "Erratum:",
}

@article{neutrino_oscillations_sk,
  author        = "{Super-Kamiokande Collaboration}",
  title         = {{Atmospheric neutrino oscillation analysis with external constraints in Super-Kamiokande I-IV}},
  journal       = {Phys. Rev. D},
  year          = 2018,
  month         = apr,
  volume        = {97},
  number        = {7},
  pages = {072001},  
  doi           = {10.1103/PhysRevD.97.072001},
  archiveprefix = {arXiv},
  eprint        = {1710.09126} 
}

@article{increased_tau_coupling,
    author = "de Gouv{\^e}a, Andr{\'e} and Vogel, Petr",
    title = "{Lepton Flavor and Number Conservation, and Physics Beyond the Standard Model}",
    eprint = "1303.4097",
    archivePrefix = "arXiv",
    primaryClass = "hep-ph",
    doi = "10.1016/j.ppnp.2013.03.006",
    journal = "Prog. Part. Nucl. Phys.",
    volume = "71",
    pages = "75--92",
    year = "2013"
}

@article{t1_seesaw_clfv_limits,
    author = "Dinh, D. N. and Ibarra, A. and Molinaro, E. and Petcov, S. T.",
    title = "{The $\mu - e$ conversion in nuclei, $\mu \to e \gamma, \mu \to 3e$ decays and TeV scale see-saw scenarios of neutrino mass generation}",
    eprint = "1205.4671",
    archivePrefix = "arXiv",
    primaryClass = "hep-ph",
    reportNumber = "FLAVOUR(267104)-ERC-15, SISSA-10-2012-EP, TUM-HEP-837-12, CFTP-12-007",
    doi = "10.1007/JHEP08(2012)125",
    journal = "JHEP",
    volume = "08",
    pages = "125",
    year = "2012",
    related        = "t1_seesaw_clfv_limits-err",
    relatedstring  = "Erratum:",
}

@article{t1_seesaw_clfv_limits2,
    author = "Alonso, R. and Dhen, M. and Gavela, M. B. and Hambye, T.",
    title = "{Muon conversion to electron in nuclei in type-I seesaw models}",
    eprint = "1209.2679",
    archivePrefix = "arXiv",
    primaryClass = "hep-ph",
    reportNumber = "ULB-TH-12-12, FTUAM-12-100, IFT-UAM-CSIC-12-78",
    doi = "10.1007/JHEP01(2013)118",
    journal = "JHEP",
    volume = "01",
    pages = "118",
    year = "2013"
}

@article{z_prime_limits,
    author = "Feruglio, Ferruccio and Paradisi, Paride and Pattori, Andrea",
    title = "{On the Importance of Electroweak Corrections for B Anomalies}",
    eprint = "1705.00929",
    archivePrefix = "arXiv",
    primaryClass = "hep-ph",
    doi = "10.1007/JHEP09(2017)061",
    journal = "JHEP",
    volume = "09",
    pages = "061",
    year = "2017"
}

@article{susy_limit,
    author = "Kajiyama, Yuji",
    title = "{Lepton flavor violating tau decay by R-parity violation in a family symmetry}",
    eprint = "0711.0286",
    archivePrefix = "arXiv",
    primaryClass = "hep-ph",
    doi = "10.24532/soken.116.1_A164",
    journal = "Soryushiron Kenkyu Electron.",
    volume = "116",
    pages = "A164--A166",
    year = "2008"
}

@article{thdm_limit,
    author = "Athron, Peter and Balazs, Csaba and Gonzalo, Tom{\'a}s E. and Jacob, Douglas and Mahmoudi, Farvah and Sierra, Cristian",
    title = "{Likelihood analysis of the flavour anomalies and g {\textendash} 2 in the general two Higgs doublet model}",
    eprint = "2111.10464",
    archivePrefix = "arXiv",
    primaryClass = "hep-ph",
    reportNumber = "CERN-TH-2021-194, TTK-21-47",
    doi = "10.1007/JHEP01(2022)037",
    journal = "JHEP",
    volume = "01",
    pages = "037",
    year = "2022"
}

@article{lhcb_run2,
    author = "{LHCb Collaboration}",
    collaboration = "LHCb",
    title = "{Search for $\tau^- \to \mu^-\mu^+\mu^-$ decays at the LHCb experiment with Run 2 data}",
    eprint = "2601.20785",
    archivePrefix = "arXiv",
    primaryClass = "hep-ex",
    reportNumber = "LHCb-PAPER-2025-052, CERN-EP-2026-003",
    month = "1",
    year = "2026"
}

@article{megii_muegamma,
    author = "{MEG II Collaboration}",
    collaboration = "MEG II",
    title = "{A search for $\mu^+ \rightarrow \textrm{e}^+ \gamma $ with the first dataset of the MEG~II experiment}",
    eprint = "2310.12614",
    archivePrefix = "arXiv",
    primaryClass = "hep-ex",
    doi = "10.1140/epjc/s10052-024-12416-2",
    journal = "Eur. Phys. J. C",
    volume = "84",
    number = "3",
    pages = "216",
    year = "2024",
    related        = "megii_muegamma-err",
    relatedstring  = "Erratum:",
}

@booklet{sv,
      author         = "Kostyukhin, V.",
      title          = "{VKalVrt - package for vertex reconstruction in ATLAS}",
      year           = "2003",
      reportNumber   = "ATL-PHYS-2003-031, ATL-COM-PHYS-2003-033,
                        CERN-ATL-PHYS-2003-031",
      report   = "ATL-PHYS-2003-031",
      howPublished   = "{ATL-PHYS-2003-031}",
      SLACcitation   = "%%CITATION = ATL-PHYS-2003-031 ETC.;%%",
      url = "http://cds.cern.ch/record/685551"
}

@article{xgboost,  
   author="Chen, Tianqi and Guestrin, Carlos",
   title={XGBoost: A Scalable Tree Boosting System},
   adsurl={http://dx.doi.org/10.1145/2939672.2939785},
   doi={10.1145/2939672.2939785},
   journal={Proceedings of the 22nd ACM SIGKDD International Conference on Knowledge Discovery and Data Mining}, 
   year={2016},
   month=aug 
}

@article{decision_trees,
    author = "Y. Coadou",
    title = "{Boosted decision trees}",
    eprint = "2206.09645",
    archivePrefix = "arXiv",
    primaryClass = "physics.data-an",
    doi = "10.1142/9789811234033_0002",
    month = "3",
    year = "2022"
}

@article{crystalball,
      author         = "Oreglia, M.",
      title          = "{A Study of the Reactions $\psi^\prime \to \gamma \gamma \psi$}",
      journal        = "SLAC-R-0236",
      url            = "www.slac.stanford.edu/cgi-wrap/getdoc/slac-r-236.pdf",
      year           = "1980",
      reportNumber   = "SLAC-0236, UMI-81-08973, SLAC-R-0236, SLAC-R-236",
      SLACcitation   = "%%CITATION = SLAC-0236 ETC.;%%",
}

@article{jet_overlap,
    author = "Cacciari, Matteo and Salam, Gavin P.",
    title = "{Pileup subtraction using jet areas}",
    eprint = "0707.1378",
    archivePrefix = "arXiv",
    primaryClass = "hep-ph",
    reportNumber = "LPTHE-07-01",
    doi = "10.1016/j.physletb.2007.09.077",
    journal = "Phys. Lett. B",
    volume = "659",
    pages = "119--126",
    year = "2008"
}

@article{ckm1,
  title = {Unitary Symmetry and Leptonic Decays},
  author = {Cabibbo, Nicola},
  journal = {Phys. Rev. Lett.},
  volume = {10},
  issue = {12},
  pages = {531},
  numpages = {0},
  year = {1963},
  month = {Jun},
  publisher = {American Physical Society},
  doi = {10.1103/PhysRevLett.10.531},
  url = "https://cds.cern.ch/record/344106"
}

@article{ckm2,
    author = {Kobayashi, Makoto and Maskawa, Toshihide},
    title = {CP-Violation in the Renormalizable Theory of Weak Interaction},
    journal = {Progress of Theoretical Physics},
    volume = {49},
    number = {2},
    pages = {652-657},
    year = {1973},
    month = {02},
    doi = {10.1143/PTP.49.652}
}


%% file: ATLAS-useful.bib
@Article{Cacciari:2008gp,
     author    = "Cacciari, Matteo and Salam, Gavin P. and Soyez, Gregory",
     title     = "{The anti-\(k_{t}\) jet clustering algorithm}",
     journal   = "JHEP",
     volume    = "04",
     year      = "2008",
     pages     = "063",
     eprint    = "0802.1189",
     archivePrefix = "arXiv",
     primaryClass  =  "hep-ph",
     doi       = "10.1088/1126-6708/2008/04/063",
     SLACcitation  = "%%CITATION = 0802.1189;%%"
}

@Article{Fastjet,
      author         = "Cacciari, Matteo and Salam, Gavin P. and Soyez, Gregory",
      title          = "{FastJet user manual}",
      journal        = "Eur. Phys. J. C",
      volume         = "72",
      year           = "2012",
      pages          = "1896",
      doi            = "10.1140/epjc/s10052-012-1896-2",
      eprint         = "1111.6097",
      archivePrefix  = "arXiv",
      primaryClass   = "hep-ph",
      reportNumber   = "CERN-PH-TH-2011-297",
      SLACcitation   = "%%CITATION = ARXIV:1111.6097;%%"
}

@Article{Pumplin:2002vw,
     author    = "Pumplin, J. and others",
     title     = "{New Generation of Parton Distributions with Uncertainties from Global QCD Analysis}",
     journal   = "JHEP",
     volume    = "07",
     year      = "2002",
     pages     = "012",
     doi       = "10.1088/1126-6708/2002/07/012",
     eprint    = "hep-ph/0201195",
     archivePrefix = "arXiv",
     SLACcitation  = "%%CITATION = HEP-PH/0201195;%%"
}

@article{Lai:2010vv,
      author         = "Lai, H.-L. and others",
      title          = "{New parton distributions for collider physics}",
      journal        = "Phys. Rev. D",
      volume         = "82",
      pages          = "074024",
      doi            = "10.1103/PhysRevD.82.074024",
      year           = "2010",
      eprint         = "1007.2241",
      archivePrefix  = "arXiv",
      primaryClass   = "hep-ph",
      reportNumber   = "MSUHEP-100707, SMU-HEP-10-10",
      SLACcitation   = "%%CITATION = ARXIV:1007.2241;%%",
}

@article{Ball:2012cx,
      author         = "{NNPDF Collaboration} and Ball, Richard D. and others",
      title          = "{Parton distributions with LHC data}",
      journal        = "Nucl. Phys. B",
      volume         = "867",
      year           = "2013",
      pages          = "244",
      doi            = "10.1016/j.nuclphysb.2012.10.003",
      eprint         = "1207.1303",
      archivePrefix  = "arXiv",
      primaryClass   = "hep-ph",
      reportNumber   = "EDINBURGH-2012-08, IFUM-FT-997, FR-PHENO-2012-014,
                        RWTH-TTK-12-25, CERN-PH-TH-2012-037, SFB-CPP-12-47\,
                        --CERN-PH-TH-2012-037",
      SLACcitation   = "%%CITATION = ARXIV:1207.1303;%%"
}

@article{Ball:2014uwa,
      author         = "{NNPDF Collaboration} and Ball, Richard D. and others",
      title          = "{Parton distributions for the LHC run II}",
      journal        = "JHEP",
      volume         = "04",
      year           = "2015",
      pages          = "040",
      doi            = "10.1007/JHEP04(2015)040",
      eprint         = "1410.8849",
      archivePrefix  = "arXiv",
      primaryClass   = "hep-ph",
      reportNumber   = "EDINBURGH-2014-15, IFUM-1034-FT, CERN-PH-TH-2013-253,
                        OUTP-14-11P, CAVENDISH-HEP-14-11",
      SLACcitation   = "%%CITATION = ARXIV:1410.8849;%%"
}

@Article{Sjostrand:2007gs,
	author    = "Sj{\"o}strand, T. and Mrenna, S. and Skands, P.",
	title     = "{A brief introduction to PYTHIA 8.1}",
	journal   = "Comput. Phys. Commun.",
	volume    = "178",
	year      = "2008",
	pages     = "852-867",
	eprint    = "0710.3820",
	archivePrefix = "arXiv",
	primaryClass  =  "hep-ph",
	doi       = "10.1016/j.cpc.2008.01.036",
	SLACcitation  = "%%CITATION = 0710.3820;%%"
}

@article{Sjostrand:2014zea,
      author         = "Sj{\"o}strand, Torbj{\"o}rn and Ask, Stefan and Christiansen,
                        Jesper R. and Corke, Richard and Desai, Nishita and Ilten,
                        Philip and Mrenna, Stephen and Prestel, Stefan and
                        Rasmussen, Christine O. and Skands, Peter Z.",
      title          = "{An introduction to PYTHIA 8.2}",
      journal        = "Comput. Phys. Commun.",
      volume         = "191",
      year           = "2015",
      pages          = "159",
      doi            = "10.1016/j.cpc.2015.01.024",
      eprint         = "1410.3012",
      archivePrefix  = "arXiv",
      primaryClass   = "hep-ph",
      reportNumber   = "LU-TP-14-36, MCNET-14-22, CERN-PH-TH-2014-190,
                        FERMILAB-PUB-14-316-CD, DESY-14-178, SLAC-PUB-16122,
                        --FERMILAB-PUB-14-316-CD",
      SLACcitation   = "%%CITATION = ARXIV:1410.3012;%%"
}

@Article{Lange:2001uf,
      author         = "Lange, D. J.",
      title          = "{The EvtGen particle decay simulation package}",
      booktitle      = "{Proceedings, 7th International Conference on B physics
                        at hadron machines (BEAUTY 2000)}",
      journal        = "Nucl. Instrum. Meth. A",
      volume         = "462",
      year           = "2001",
      pages          = "152",
      doi            = "10.1016/S0168-9002(01)00089-4",
      SLACcitation   = "%%CITATION = NUIMA,A462,152;%%"
}

@Article{Nason:2004rx,
      author         = "Nason, Paolo",
      title          = "{A new method for combining NLO QCD with shower Monte Carlo algorithms}",
      journal        = "JHEP",
      volume         = "11",
      pages          = "040",
      doi            = "10.1088/1126-6708/2004/11/040",
      year           = "2004",
      eprint         = "hep-ph/0409146",
      archivePrefix  = "arXiv",
}

@Article{Frixione:2007vw,
      author         = "Frixione, Stefano and Nason, Paolo and Oleari, Carlo",
      title          = "{Matching NLO QCD computations with parton shower
                        simulations: the POWHEG method}",
      journal        = "JHEP",
      volume         = "11",
      pages          = "070",
      doi            = "10.1088/1126-6708/2007/11/070",
      year           = "2007",
      eprint         = "0709.2092",
      archivePrefix  = "arXiv",
      primaryClass   = "hep-ph",
}

@Article{Alioli:2010xd,
      author         = "Alioli, Simone and Nason, Paolo and Oleari, Carlo and Re,
                        Emanuele",
      title          = "{A general framework for implementing NLO calculations in
                        shower Monte Carlo programs: the POWHEG BOX}",
      journal        = "JHEP",
      volume         = "06",
      pages          = "043",
      doi            = "10.1007/JHEP06(2010)043",
      year           = "2010",
      eprint         = "1002.2581",
      archivePrefix  = "arXiv",
      primaryClass   = "hep-ph",
}

@article{Alioli:2008gx,
      author         = "Alioli, Simone and Nason, Paolo and Oleari, Carlo and Re,
                        Emanuele",
      title          = "{NLO vector-boson production matched with shower in
                        POWHEG}",
      journal        = "JHEP",
      volume         = "07",
      year           = "2008",
      pages          = "060",
      doi            = "10.1088/1126-6708/2008/07/060",
      eprint         = "0805.4802",
      archivePrefix  = "arXiv",
      primaryClass   = "hep-ph",
      reportNumber   = "NSF-KITP-08-74",
      SLACcitation   = "%%CITATION = ARXIV:0805.4802;%%"
}

@article{Frixione:2007nw,
      author         = "Frixione, Stefano and Ridolfi, Giovanni and Nason, Paolo",
      title          = "{A positive-weight next-to-leading-order Monte Carlo for
                        heavy flavour hadroproduction}",
      journal        = "JHEP",
      volume         = "09",
      pages          = "126",
      doi            = "10.1088/1126-6708/2007/09/126",
      year           = "2007",
      eprint         = "0707.3088",
      archivePrefix  = "arXiv",
      primaryClass   = "hep-ph",
      SLACcitation   = "%%CITATION = ARXIV:0707.3088;%%",
}

@article{Agostinelli:2002hh,
      author         = "Agostinelli, S. and others",
      title          = "{\textsc{Geant4} -- a simulation toolkit}",
      journal        = "Nucl. Instrum. Meth. A",
      volume         = "506",
      year           = "2003",
      pages          = "250",
      doi            = "10.1016/S0168-9002(03)01368-8",
      reportNumber   = "SLAC-PUB-9350, FERMILAB-PUB-03-339",
      SLACcitation   = "%%CITATION = NUIMA,A506,250;%%"
}

@Article{Read:2002hq,
    Author = {Read, Alexander L.},
    Journal = {J. Phys. G},
    Pages = {2693},
    doi = "10.1088/0954-3899/28/10/313",
    Title = "{Presentation of search results: the \(CL_s\) technique}",
    Volume = {28},
    Year = {2002}
}

@Article{Cowan:2010js,
   Author = {Cowan, Glen and Cranmer, Kyle and Gross, Eilam and Vitells, Ofer},
   Title = "{Asymptotic formulae for likelihood-based tests of new physics}",
   Journal = {Eur. Phys. J. C},
   Volume = {71},
   Year = {2011},
   Pages = {1554},
   doi = "10.1140/epjc/s10052-011-1554-0",
   Eprint = "1007.1727",
   Archiveprefix = {arXiv},
   Primaryclass = {physics.data-an},
   related = "Cowan:2010js-err",
   relatedstring  = "Erratum:",
}

@article{Zyla:2020zbs,
    author = "{Particle Data Group} and Zyla, P.A. and others",
    title = "{Review of Particle Physics}",
    journal = "Prog. Theor. Exp. Phys.",
    volume = "2020",
    number = "8",
    pages = "083C01",
    year = "2020",
    doi = "10.1093/ptep/ptaa104",
}


%% file: ATLAS.bib
@Article{PERF-2007-01,
    author         = "{ATLAS Collaboration}",
    title          = "{The ATLAS Experiment at the CERN Large Hadron Collider}",
    journal        = "JINST",
    volume         = "3",
    year           = "2008",
    pages          = "S08003",
    doi            = "10.1088/1748-0221/3/08/S08003",
    primaryClass   = "hep-ex",
}

@Article{SOFT-2010-01,
    author         = "{ATLAS Collaboration}",
    title          = "{The ATLAS Simulation Infrastructure}",
    journal        = "Eur. Phys. J. C",
    volume         = "70",
    year           = "2010",
    pages          = "823",
    doi            = "10.1140/epjc/s10052-010-1429-9",
    eprint         = "1005.4568",
    archivePrefix  = "arXiv",
    primaryClass   = "physics.ins-det",
}

@Article{STDM-2012-23,
    author         = "{ATLAS Collaboration}",
    title          = "{Measurement of the \(Z/\gamma^*\) boson transverse momentum distribution in \(pp\) collisions at \(\sqrt{s} = 7\,\text{TeV}\) with the ATLAS detector}",
    journal        = "JHEP",
    volume         = "09",
    year           = "2014",
    pages          = "145",
    doi            = "10.1007/JHEP09(2014)145",
    reportNumber   = "CERN-PH-EP-2014-075",
    eprint         = "1406.3660",
    archivePrefix  = "arXiv",
    primaryClass   = "hep-ex",
}

@Article{HIGG-2013-08,
    author         = "{ATLAS Collaboration}",
    title          = "{Measurement of Higgs boson production in the diphoton decay channel in \(pp\) collisions at center-of-mass energies of \(7\) and \(8\,\text{TeV}\) with the ATLAS detector}",
    journal        = "Phys. Rev. D",
    volume         = "90",
    year           = "2014",
    pages          = "112015",
    doi            = "10.1103/PhysRevD.90.112015",
    reportNumber   = "CERN-PH-EP-2014-198",
    eprint         = "1408.7084",
    archivePrefix  = "arXiv",
    primaryClass   = "hep-ex",
}

@Article{EXOT-2014-14,
    author         = "{ATLAS Collaboration}",
    title          = "{Probing lepton flavour violation via neutrinoless \(\tau\rightarrow 3\mu\) decays with the ATLAS detector}",
    journal        = "Eur. Phys. J. C",
    volume         = "76",
    year           = "2016",
    pages          = "232",
    doi            = "10.1140/epjc/s10052-016-4041-9",
    reportNumber   = "CERN-PH-EP-2015-315",
    eprint         = "1601.03567",
    archivePrefix  = "arXiv",
    primaryClass   = "hep-ex",
}

@Article{HIGG-2014-04,
    author         = "{ATLAS Collaboration}",
    title          = "{Search for Scalar Diphoton Resonances in the Mass Range 65--600\,GeV with the ATLAS Detector in \(pp\) Collision Data at \(\sqrt{s} = 8\,\text{TeV}\)}",
    journal        = "Phys. Rev. Lett.",
    volume         = "113",
    year           = "2014",
    pages          = "171801",
    doi            = "10.1103/PhysRevLett.113.171801",
    reportNumber   = "CERN-PH-EP-2014-142",
    eprint         = "1407.6583",
    archivePrefix  = "arXiv",
    primaryClass   = "hep-ex",
}

@Article{PERF-2015-09,
    author         = "{ATLAS Collaboration}",
    title          = "{Jet reconstruction and performance using particle flow with the ATLAS Detector}",
    journal        = "Eur. Phys. J. C",
    volume         = "77",
    year           = "2017",
    pages          = "466",
    doi            = "10.1140/epjc/s10052-017-5031-2",
    reportNumber   = "CERN-EP-2017-024",
    eprint         = "1703.10485",
    archivePrefix  = "arXiv",
    primaryClass   = "hep-ex",
}

@Article{STDM-2015-05,
    author         = "{ATLAS Collaboration}",
    title          = "{Measurement of the Inelastic Proton--Proton Cross Section at \(\sqrt{s} = 13\,\text{TeV}\) with the ATLAS Detector at the LHC}",
    journal        = "Phys. Rev. Lett.",
    volume         = "117",
    year           = "2016",
    pages          = "182002",
    doi            = "10.1103/PhysRevLett.117.182002",
    reportNumber   = "CERN-EP-2016-140",
    eprint         = "1606.02625",
    archivePrefix  = "arXiv",
    primaryClass   = "hep-ex",
}

@Article{TRIG-2016-01,
    author         = "{ATLAS Collaboration}",
    title          = "{Performance of the ATLAS trigger system in 2015}",
    journal        = "Eur. Phys. J. C",
    volume         = "77",
    year           = "2017",
    pages          = "317",
    doi            = "10.1140/epjc/s10052-017-4852-3",
    reportNumber   = "CERN-EP-2016-241",
    eprint         = "1611.09661",
    archivePrefix  = "arXiv",
    primaryClass   = "hep-ex",
}

@Article{BPHY-2018-05,
    author         = "{ATLAS Collaboration}",
    title          = "{A search for lepton-flavour violating \(\tau \rightarrow 3\mu\) decays with the ATLAS detector}",
    year           = "2026",
    reportNumber   = "CERN-EP-2026-059",
    eprint         = "2603.18099",
    archivePrefix  = "arXiv",
    primaryClass   = "hep-ex",
}

@Article{BPHY-2018-10,
    author         = "{ATLAS Collaboration}",
    title          = "{Differential cross-section measurements of \(D^{\pm}\) and \(D_{s}^{\pm}\) meson production in proton--proton collisions at \(\sqrt{s} = 13\,\text{TeV}\) with the ATLAS detector}",
    journal        = "JHEP",
    volume         = "07",
    year           = "2025",
    pages          = "086",
    doi            = "10.1007/JHEP07(2025)086",
    reportNumber   = "CERN-EP-2024-322",
    eprint         = "2412.15742",
    archivePrefix  = "arXiv",
    primaryClass   = "hep-ex",
}

@Article{DAPR-2018-01,
    author         = "{ATLAS Collaboration}",
    title          = "{ATLAS data quality operations and performance for 2015--2018 data-taking}",
    journal        = "JINST",
    volume         = "15",
    year           = "2020",
    pages          = "P04003",
    doi            = "10.1088/1748-0221/15/04/P04003",
    reportNumber   = "CERN-EP-2019-207",
    eprint         = "1911.04632",
    archivePrefix  = "arXiv",
    primaryClass   = "physics.ins-det",
}

@Article{JETM-2018-05,
    author         = "{ATLAS Collaboration}",
    title          = "{Jet energy scale and resolution measured in proton--proton collisions at \(\sqrt{s} = 13\,\text{TeV}\) with the ATLAS detector}",
    journal        = "Eur. Phys. J. C",
    volume         = "81",
    year           = "2021",
    pages          = "689",
    doi            = "10.1140/epjc/s10052-021-09402-3",
    reportNumber   = "CERN-EP-2020-083",
    eprint         = "2007.02645",
    archivePrefix  = "arXiv",
    primaryClass   = "hep-ex",
}

@Article{MUON-2018-03,
    author         = "{ATLAS Collaboration}",
    title          = "{Muon reconstruction and identification efficiency in ATLAS using the full Run~2 \(pp\) collision data set at \(\sqrt{s} = 13\,\text{TeV}\)}",
    journal        = "Eur. Phys. J. C",
    volume         = "81",
    year           = "2021",
    pages          = "578",
    doi            = "10.1140/epjc/s10052-021-09233-2",
    reportNumber   = "CERN-EP-2020-199",
    eprint         = "2012.00578",
    archivePrefix  = "arXiv",
    primaryClass   = "hep-ex",
}

@Article{STDM-2018-17,
    author         = "{ATLAS Collaboration}",
    title          = "{Precise measurements of \(W\)- and \(Z\)-boson transverse momentum spectra with the ATLAS detector using \(pp\) collisions at \(\sqrt{s} = 5.02\,\text{TeV}\) and \(13\,\text{TeV}\)}",
    journal        = "Eur. Phys. J. C",
    volume         = "84",
    year           = "2024",
    pages          = "1126",
    doi            = "10.1140/epjc/s10052-024-13414-0",
    reportNumber   = "CERN-EP-2024-080",
    eprint         = "2404.06204",
    archivePrefix  = "arXiv",
    primaryClass   = "hep-ex",
}

@Article{TOPQ-2018-26,
    author         = "{ATLAS Collaboration}",
    title          = "{Inclusive and differential cross-sections for dilepton \(t\bar{t}\) production measured in \(\sqrt{s} = 13\,\text{TeV}\) \(pp\) collisions with the ATLAS detector}",
    journal        = "JHEP",
    volume         = "07",
    year           = "2023",
    pages          = "141",
    doi            = "10.1007/JHEP07(2023)141",
    reportNumber   = "CERN-EP-2023-016",
    eprint         = "2303.15340",
    archivePrefix  = "arXiv",
    primaryClass   = "hep-ex",
}

@Article{TRIG-2018-01,
    author         = "{ATLAS Collaboration}",
    title          = "{Performance of the ATLAS muon triggers in Run~2}",
    journal        = "JINST",
    volume         = "15",
    year           = "2020",
    pages          = "P09015",
    doi            = "10.1088/1748-0221/15/09/p09015",
    reportNumber   = "CERN-EP-2020-031",
    eprint         = "2004.13447",
    archivePrefix  = "arXiv",
    primaryClass   = "physics.ins-det",
}

@Article{JETM-2020-03,
    author         = "{ATLAS Collaboration}",
    title          = "{The performance of missing transverse momentum reconstruction and its significance with the ATLAS detector using \(140\,\text{fb}^{-1}\) of \(\sqrt{s} = 13\,\text{TeV}\) \(pp\) collisions}",
    journal        = "Eur. Phys. J. C",
    volume         = "85",
    year           = "2025",
    pages          = "606",
    doi            = "10.1140/epjc/s10052-025-14062-8",
    reportNumber   = "CERN-EP-2024-023",
    eprint         = "2402.05858",
    archivePrefix  = "arXiv",
    primaryClass   = "hep-ex",
}

@Article{DAPR-2021-01,
    author         = "{ATLAS Collaboration}",
    title          = "{Luminosity determination in \(pp\) collisions at \(\sqrt{s} = 13\,\text{TeV}\) using the ATLAS detector at the LHC}",
    journal        = "Eur. Phys. J. C",
    volume         = "83",
    year           = "2023",
    pages          = "982",
    doi            = "10.1140/epjc/s10052-023-11747-w",
    reportNumber   = "CERN-EP-2022-281",
    eprint         = "2212.09379",
    archivePrefix  = "arXiv",
    primaryClass   = "hep-ex",
}

@Article{MUON-2022-01,
    author         = "{ATLAS Collaboration}",
    title          = "{Studies of the muon momentum calibration and performance of the ATLAS detector with \(pp\) collisions at \(\sqrt{s} = 13\,\text{TeV}\)}",
    journal        = "Eur. Phys. J. C",
    volume         = "83",
    year           = "2023",
    pages          = "686",
    doi            = "10.1140/epjc/s10052-023-11584-x",
    reportNumber   = "CERN-EP-2022-274",
    eprint         = "2212.07338",
    archivePrefix  = "arXiv",
    primaryClass   = "hep-ex",
}

@Article{SOFT-2022-02,
    author         = "{ATLAS Collaboration}",
    title          = "{Software and computing for Run~3 of the ATLAS experiment at the LHC}",
    journal        = "Eur. Phys. J. C",
    volume         = "85",
    year           = "2025",
    pages          = "234",
    doi            = "10.1140/epjc/s10052-024-13701-w",
    reportNumber   = "CERN-EP-2024-100",
    eprint         = "2404.06335",
    archivePrefix  = "arXiv",
    primaryClass   = "hep-ex",
    related        = "SOFT-2022-02-err",
    relatedstring  = "Erratum:",
}

@Booklet{ATL-SOFT-PUB-2026-001,
    author         = "{ATLAS Collaboration}",
    title          = "{ATLAS Computing Acknowledgements}",
    howpublished   = "{ATL-SOFT-PUB-2026-001}",
    url            = "https://cds.cern.ch/record/2952666",
    year           = "2026",
}

@Article{LUCID2,
    author         = "Avoni, G. and others",
    title          = "{The new LUCID-2 detector for luminosity measurement and monitoring in ATLAS}",
    journal        = "JINST",
    volume         = "13",
    year           = "2018",
    pages          = "P07017",
    doi            = "10.1088/1748-0221/13/07/P07017",
}


%% file: CMS.bib
@Article{CMS-BPH-17-004,
    author         = "{CMS Collaboration}",
    title          = "{Search for the lepton flavor violating decay \(\tau \rightarrow 3\mu\) in proton--proton collisions at \(\sqrt{s} = 13\,\text{TeV}\)}",
    journal        = "JHEP",
    volume         = "01",
    year           = "2021",
    pages          = "163",
    doi            = "10.1007/JHEP01(2021)163",
    reportNumber   = "CERN-EP-2020-113",
    eprint         = "2007.05658",
    archivePrefix  = "arXiv",
    primaryClass   = "hep-ex",
}

@Article{CMS-BPH-21-005,
    author         = "{CMS Collaboration}",
    title          = "{Search for the lepton flavor violating \(\tau \rightarrow 3\mu\) decay in proton--proton collisions at \(\sqrt{s} = 13\,\text{TeV}\)}",
    journal        = "Phys. Lett. B",
    volume         = "853",
    year           = "2024",
    pages          = "138633",
    doi            = "10.1016/j.physletb.2024.138633",
    reportNumber   = "CERN-EP-2023-249",
    eprint         = "2312.02371",
    archivePrefix  = "arXiv",
    primaryClass   = "hep-ex",
}


%% file: PubNotes.bib
@Booklet{ATL-PHYS-PUB-2014-021,
    author         = "{ATLAS Collaboration}",
    title          = "{ATLAS Pythia~8 tunes to \(7\,\text{TeV}\) data}",
    howpublished   = "{ATL-PHYS-PUB-2014-021}",
    url            = "https://cds.cern.ch/record/1966419",
    year           = "2014",
}

@Booklet{ATL-PHYS-PUB-2016-017,
    author         = "{ATLAS Collaboration}",
    title          = "{The Pythia~8 A3 tune description of ATLAS minimum bias and inelastic measurements incorporating the Donnachie--Landshoff diffractive model}",
    howpublished   = "{ATL-PHYS-PUB-2016-017}",
    url            = "https://cds.cern.ch/record/2206965",
    year           = "2016",
}

@Booklet{ATL-PHYS-PUB-2020-028,
    author         = "{ATLAS Collaboration}",
    title          = "{Recommendations for the Modeling of Smooth Backgrounds}",
    howpublished   = "{ATL-PHYS-PUB-2020-028}",
    url            = "https://cds.cern.ch/record/2743717",
    year           = "2020",
}
